# Sliding Over Graphene Grain Boundaries: A Step Towards Macroscale Superlubricity

Xiang Gao, Wengen Ouyang, Oded Hod,[*] and Michael Urbakh

Department of Physical Chemistry, School of Chemistry, The Raymond and Beverly Sackler Faculty of Exact Sciences and The Sackler Center for Computational Molecular and Materials Science, Tel Aviv University, Tel Aviv 6997801, Israel.

**Abstract**

In light of the race towards macroscale superlubricity of graphitic contacts, the effect of grain boundaries on their frictional properties becomes of central importance. Here, we elucidate the unique frictional mechanisms characterizing topological defects along typical grain boundaries that can vary from being nearly flat to highly corrugated, depending on the boundary misfit angle. We find that frictional energy dissipation over grain boundaries can originate from variations of compressibility along the surface, heat produced during defect (un)buckling events, and elastic energy storage in irreversible buckling processes. These may lead to atypical non-monotonic dependence of the averaged friction on the normal load. The knowledge gained in the present study constitutes an important step towards the realization of superlubricity in macroscopic graphitic contacts.

[*]Corresponding author. email address: odedhod@tauex.tau.ac.il

# Introduction

Structural superlubricity, the intriguing behavior of ultralow friction at incommensurate solid/solid crystalline interfaces, has emerged as a promising way toward efficient reduction of energy dissipation and wear at various length scales.[1] This phenomenon has been experimentally demonstrated in various contacts of pristine single-crystal layered materials, including nano- and microscale graphitic structures,[2, 3] graphene/*h*-BN heterojunctions,[4-7] homogenous $MoS_2$ interfaces,[8-10], graphene/$MoS_2$ heterojunctions,[11] as well as other crystalline interfaces.[12] The common feature characterizing contacts of layered materials is their anisotropic structure consisting of covalently bonded layers coupled to each other via weaker dispersive interactions. A necessary condition to obtain such structural superlubricity is incommensurability between the rigid crystalline network of the contacting surfaces. For homogeneous layered materials junctions this can be fulfilled when the contacting surfaces are laterally rotated with respect to each other, thus avoiding frictional lock-in at the commensurate aligned configuration.[2] The latter state is eliminated in heterojunctions that possess mismatching lattice periodicity, providing superlubric behavior that is robust against interface reorientations.[4-6]

The demonstration of superlubric behavior at the nano- and micro-scales has triggered scientific and engineering efforts aiming to extend its scope towards the macroscale.[13-15] Within the realm of layered materials interfaces, further scaling-up of structural superlubricity towards the macroscopic world is, nevertheless, challenged by the polycrystalline nature of layered materials at larger scales. At increasing contact dimensions, the contacting surfaces often exhibit a mosaic of randomly oriented grains that are separated by grain boundaries (GBs) in the form of chains of dislocations.[16] The randomness of the grain orientations has the potential to eliminate frictional lock-in due to reduced interfacial commensurability thus promoting superlubric behavior.[17, 18] However, the GBs typically induce large out-of-plane surface deformations[19-21] that may degrade or even completely eliminate the superlubric behavior, as well as impact wear resistance.[22] Moreover, if the polycrystalline surface grain density is relatively uniform, the overall GB length will scale with surface area and hence their frictional effects are expected to scale with surface area as well, thus eliminating the sublinear scaling of superlubricity with contact surface area.[23] Therefore, understanding the effect of GBs on the frictional behavior of layered-materials contacts is imperative for achieving superlubricity at large-scale interfaces. Prior to studying the complex frictional behavior of the entire mosaic structure,[24, 25] it is instructive to focus first on the effects of individual topological defects that constitute a single GB. This allows us to elucidate the mechanisms of energy dissipation induced by such topological barriers and identify the corresponding characteristic frictional signatures.

To this end, we adopt a model system of polycrystalline graphene[20, 21, 26-29] with a single GB that consists of a chain of pentagon-heptagon pair dislocations (see Figure 1a). To study the localized effect of one or a few adjacent GB dislocations we consider a small pristine tri-layer graphene (TLG) flake sliding over a limited section of the entire GB (see Figure 3a). To allow for the application of an external load, the entire system is placed on a pristine Bernal (AB)-stacked bilayer graphene (BLG) substrate, whose lower layer is kept rigid. We first study the structural properties of GBs for various relative orientations of the contacting grains using a dedicated classical force-field.[30-34] Then, we perform fully-atomistic molecular dynamics simulations to study the frictional properties of the flake when sliding over individual GB dislocations under various normal loads and identify the mechanisms underlying their tribological behavior. Details regarding the structural relaxation and dynamic simulations are given in the Methods section.

**Structural Properties of Graphene Grain Boundaries**

Graphene GBs typically consist of an array of pentagon-heptagon dislocations (see Figure 1a), whose exact configuration is dictated by the relative orientation of the contacting grains (see Supporting Information (SI) section 1 for further details). These topological defects lead to high local stress fields that can be relieved via out-of-plane surface deformation (see Figure 1b and SI section 2). Since the substrate limits the possibility of downward protrusion, the local stresses developing at the GB are relieved via upward out-of-plan deformation. Such deformations have been experimentally characterized with both scanning tunneling microscopy (STM) and atomic force microscopy (AFM).[20, 21] Notably, the measured corrugation obtained by STM (~8-15 Å) is considerably larger than that measured by AFM (~3 Å). Hence, to better understand the relation between the orientation of the contacting grains and the structure of the GB, we studied a set of systems each consisting of two grains at different relative orientations. One of the grains (marked as "Grain 1") is kept at AB stacking with the underlying pristine BLG substrate, whose armchair direction is placed along $x$-axis. The orientation of the other grain (Grain 2) is varied in a counter-clockwise direction with respect to the $x$-axis to generate GBs of misfit angles $\theta$ varying from 0° to 60° (see e.g. Figure 1a). We note that due to the 6-fold symmetry of the underlying pristine graphene layer, this misfit angle range is complete. Following structural relaxation, two types of GBs are obtained: (i) Corrugated GBs that exhibit a series of upward protruded bumps along the GB, as shown in Figure 1b,c. (ii) Flat GBs obtained for misfit angles in the range $21.8° < \theta < 38.2°$ (see Figure 1d) in agreement with experimental observations.[21] Specifically, we obtain such flat GBs for the angles $\theta$=21.8°, 25.7°, 27.8°, and 30° and their symmetric reflections around the 30° line. When analyzing the dependence of the topography and energetics of the GBs on the misfit angle, we find it necessary to further divide these GB types into two subgroups each (marked as Corrugated-I, Corrugated-II, Flat-I, and Flat-II). The different misfit angle dependence of each GB subgroup stems from the specific configuration of the corresponding GB lattice dislocations (see SI section 3 for further details).

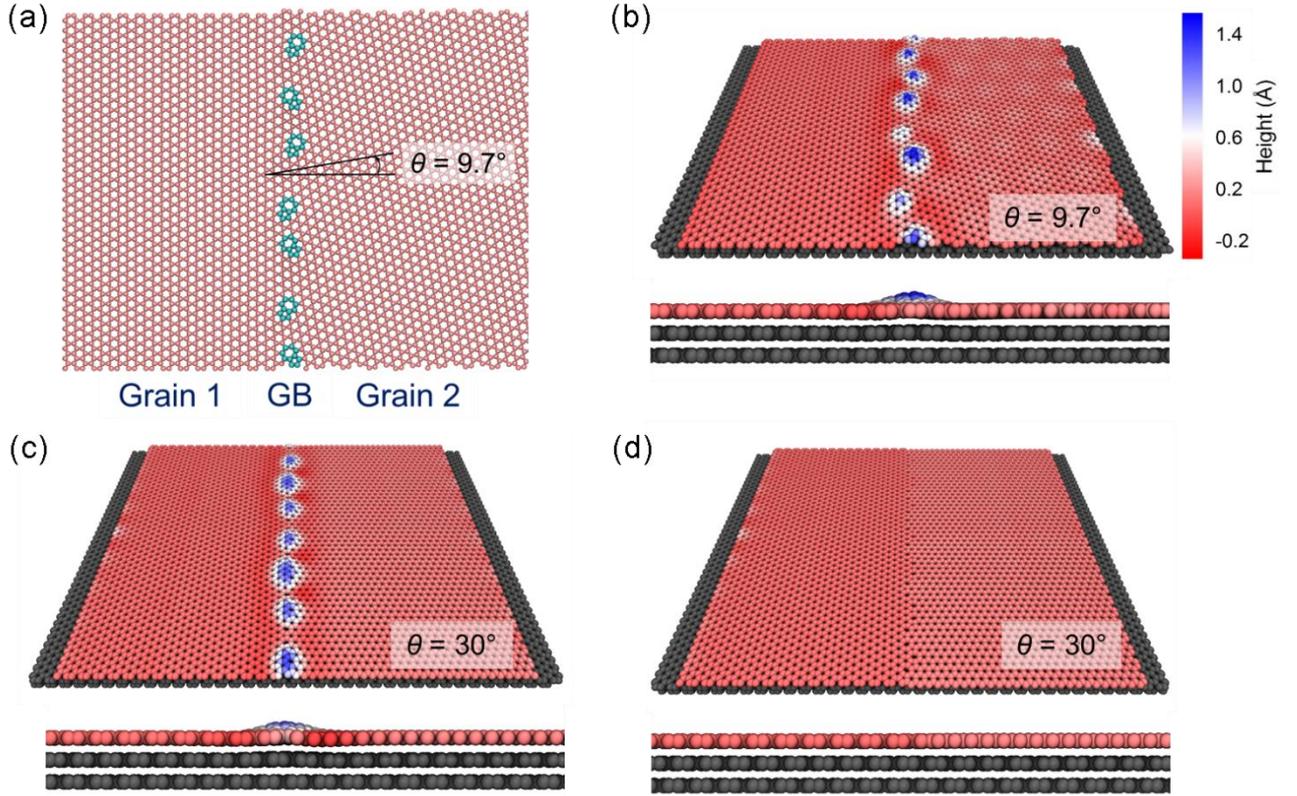

**Figure 1. Structure and topography of graphene GBs.** (a) Atomic structure of a graphene GB with a misfit angle $\theta$ = 9.7°. The cyan and pink spheres represent the pentagon-heptagon dislocation atoms and hexagonal carbon atoms, respectively. (b)-(d) Perspective and cross section views of graphene GBs on a BLG substrate with different misfit angles. (b) Corrugated GB with misfit angle $\theta$ = 9.7°, same GB as (a). (c) Corrugated GB and (d) flat GB with the same misfit angle $\theta$ = 30°. The color of the spheres in the polycrystalline graphene layer in (b)-(d) represents the atomic height with respect to the average height of the two grains, using the color scale in (b). The grey spheres represent the carbon atoms of the pristine BLG substrate. The details for generating the configuration of graphene GB can be found in the Method section.

In the range 0°< $\theta$ < 21.8° only Corrugated-I GBs appear. For small misfit angles, a bump height of ~2.2 Å is obtained (see Figure 2a) in good agreement with the experimental AFM observations.[20, 21] When increasing the misfit angle in the range 0° < $\theta$ < 12.5°, we observe a rapid increase in the dislocation density (Figure 2b) in accordance with theoretical predictions.[19, 35] This induces stress cancellation between neighboring dislocations leading to a substantial decrease of the maximum bump height (from 2.2 Å to about 1.4 Å) with misfit angle up to $\theta$ = 12.5° (see Figure 2a). The combined effect of this structural behavior results in a rapid increase in the GB energy (see Figure 2c) relative to the energy of the corresponding separated grains (See SI section 4 for further details on the GB energy calculation). In the subrange of 12.5° < $\theta$ < 21.8° the density of the Corrugated-I GB bumps generally reduces and their maximum height shows merely weak $\theta$-dependence resulting in a mild reduction of the overall GB energy. Within this range, the dislocations tend to share a C-C bond, a hexagon carbon ring, or a pair of mutual edge atoms, which significantly reduces the local stress field and annihilates bumps. Finally, the 21.8° < $\theta$ < 30° GBs form quasi- or fully continuous dislocation chains (see SI section 3) of either Corrugated-I, Corrugated-II, or flat types. The Corrugated-II GBs show weak bump density and height dependence on the misfit angle as compared to the Corrugated-I subgroup that exhibits further density and height reduction (Figure 2a,b). Notably, the maximum out-of-plane corrugation of the flat GBs of both types is below 0.2 Å, which is significantly lower than that of the corrugated counterparts and is found to be independent of the misfit angle. At this misfit angle range, the difference in geometric structure of the various GBs leads to energetic splitting, where the Corrugated-I and Flat-I subgroups show reduction of GB energy with increased misfit angle, whereas the Corrugated-II and Flat-II subgroups show larger GB energy and weak $\theta$ dependence (see Figure 2c). We note that the GB energies for the Corrugated-I and Flat-I GBs are comparable. This is because after out-of-plane deformation, the GB energy is mainly determined by the dislocation types and densities. Since the Corrugated-I and Flat-I GBs share same type and similar dislocation density, their GB energies are comparable. A similar picture arises for the Corrugated-II and Flat-II GBs (see SI section 5 for further details).

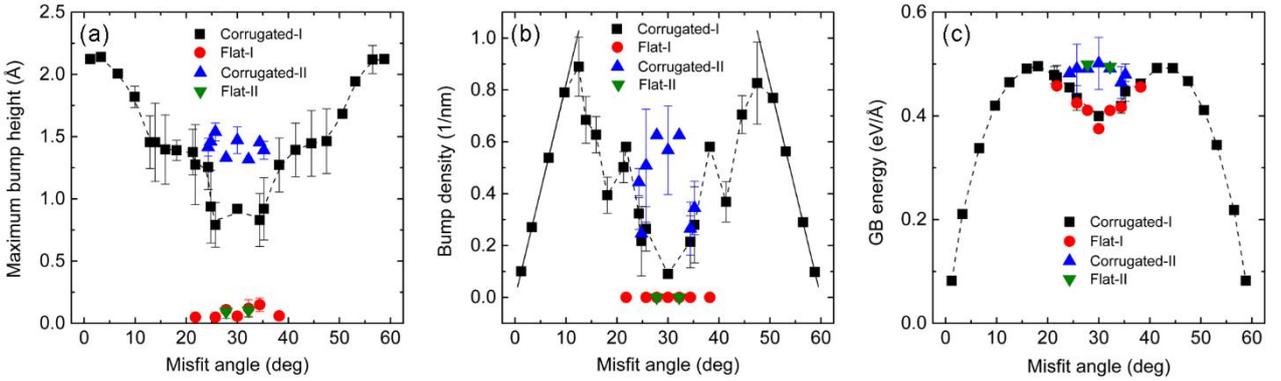

**Figure 2. Structure and energetics of graphene GBs.** (a) Maximum bump height; (b) Graphene GB bump density; and (c) GB energy plotted as a function of misfit angle $\theta$. Based on the corrugation and the types of dislocations, the GBs are categorized in four groups: Corrugated-I (black squares), Corrugated-II (blue up triangles), Flat-I (red circles), and Flat-II (green down triangles). See SI sections 1 and 3 for further details regarding the various GB types. The black solid lines in (b) indicate the theoretical prediction.[19, 35] The black dashed lines in all panels are plotted to guide the eye. For flat GBs, the bump density is presented as zero and the maximum out-of-plane corrugation is shown in (a). Details regarding the estimation of the error bars can be found in the Method section. Error bars smaller than the symbol size are not shown.

**Frictional Properties of Graphene Grain Boundaries**

From the above analysis, we can expect that the frictional properties of interfaces consisting of polycrystalline graphene surfaces will depend on the relative orientation of the various grains. To gain microscopic understanding of the tribological effect of GBs, we considered three typical GBs of misfit angles 4.7°, 13.9° and 27.8° that represent the low bump density/high corrugation case, where the flake slides over a single bump; high bump density/low corrugation case, where the flake slides over two bumps; and flat GB topologies, respectively. Panels (a) and (b) of Figure 3 show the setup for the friction simulations with the $\theta = 4.7°$ GB, where the TLG flake slides atop a bi-crystalline graphene surface with a single GB supported by the pristine BLG substrate, whose lower layer is fixed. The slider is pulled from its optimal configuration by moving its top rigid layer with a constant velocity of 5 m/s across the GB (see Method section for further details). The frictional stress acting on the slider when crossing the GB is calculated as the average lateral force in the region near the GB divided by the flake's surface area. To avoid intractable computational burden, all simulations presented below are performed using a damped dynamics algorithm, i.e. Langevin dynamics at zero temperature.[36] Exemplary simulations performed at room temperature provide similar results (see SI section 8).

To estimate the friction coefficient associated with crossing the GB, we calculate the dependence of the frictional stress on the normal load applied to the upper rigid layer of the slider. Figure 3c presents the frictional stress obtained for the three GBs considered at the forward sliding direction (from Grain 1 to Grain 2, as noted in Figure 1) compared to the pristine graphene surface results. All three representative GBs considered show an increase of the frictional stress with applied normal load, which is considerably steeper than that of the corresponding pristine surface. Notably, the two corrugated GBs ($\theta = 4.7°$, 13.9°) show a sharp jump in the frictional stress at low normal loads, which is absent in the flat GB case (27.8°). At higher loads, the frictional stresses of the three GBs behave qualitatively similar, showing a linear increase with normal load.

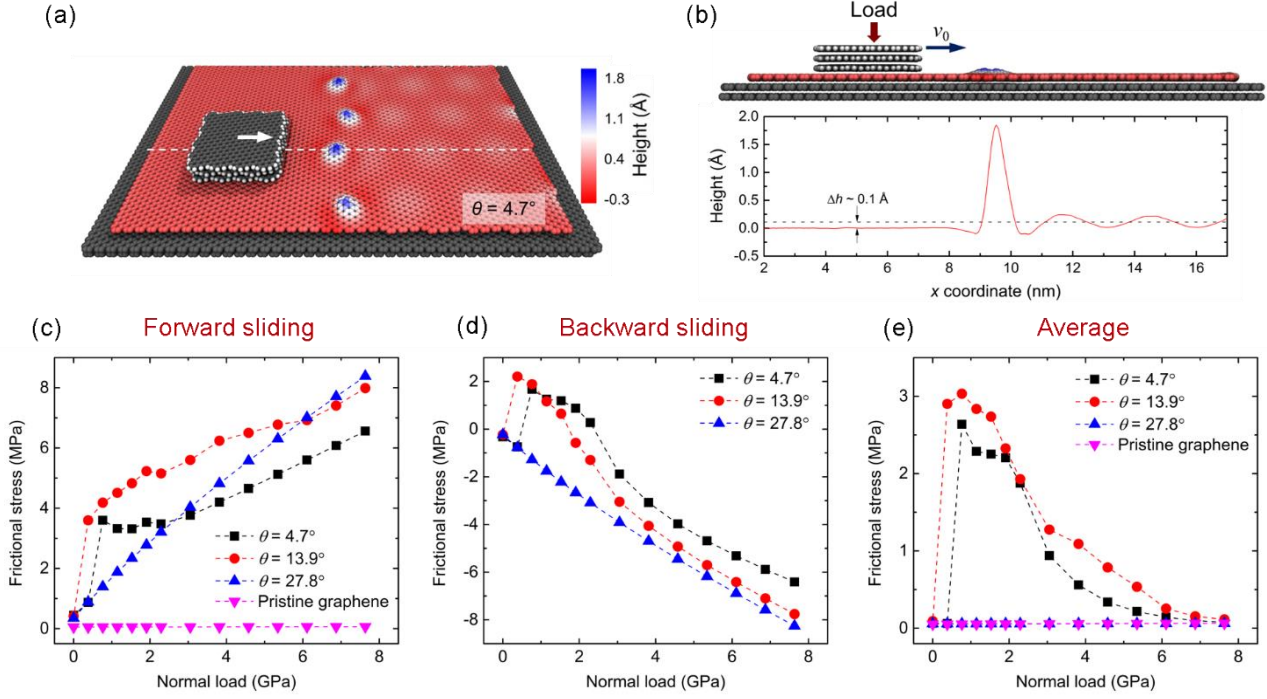

**Figure 3. Sliding simulation setup for a graphene flake over a graphene GB.** (a) Schematic representation of the simulation setup for a TLG flake sliding over a GB of misfit angle $\theta = 4.7°$. Polycrystalline substrate atoms are colored according to their vertical height. The grey and white spheres in other layers represent carbon and hydrogen atoms, respectively. The white dashed line depicts the scan line and the arrow shows the forward sliding direction. (b) Top: side view of the sliding system presented in panel (a); Bottom: the height profile along the white dashed scan line shown in panel (a) in absence of the flake. The black dashed line in (b) indicates the average height of Grain 2. (c)-(e) show the frictional stress as a function of normal load for the $\theta = 4.7°$ (black squares), 13.9° (red circles) and 27.8° (blue upper triangles) GBs in the forward and backward sliding directions and their average, respectively. Results obtained for a pristine graphene surface are presented for comparison by the magenta lower triangle symbols in panels (c) and (e). The dashed lines in (c-e) are plotted to guide the eye. More details regarding the friction calculations and the effect of scan-line position can be found in SI sections 6 and 7, respectively. Movies of typical simulations are also provided in SI.

The different behavior of the corrugated and flat GBs at the low normal load regime results from bump buckling and unbuckling processes that occur when sliding over corrugated GBs. To demonstrate this, we consider the $\theta = 4.7°$ GB and plot the corresponding lateral force traces (Figure 4a) and the variations of the bump height and its vertical velocity (Figure 4b) for three normal loads, corresponding to the different regimes discussed above. Starting at steady state over the left grain, the lateral force trace shows a resistive contribution when encountering the GB at zero normal load (see SI Movie 1). This is associated with a mild (~1 Å) depression of the bump height. During the entire crossing process, the bump remains depressed and the lateral force shows somewhat enhanced oscillations. When the slider approaches the edge of the GB, the bump recovers resulting in an assistive lateral force leading the flake towards a new steady-state sliding over Grain 2. Importantly, in this case the shear induced bump depression is nearly adiabatic as reflected by the minor variations obtained in the vertical velocity. This, in turn, results in the low frictional stress obtained for the load free case. When increasing the load to 0.8 GPa, the bump dynamics changes dramatically (see SI Movie 2). While the force trace exhibits similar trends as those of the load-free case, pronounced buckling occurs during the crossing process, where the bump is suppressed below the surface while storing an elastic energy of 0.8 eV (see SI section 9). Notably, while this buckling is associated with an energy barrier of ~0.9 eV, the corresponding unbuckling process has a significantly lower barrier of ~0.1 eV. This explains why many of our dynamical simulations exhibit reversible buckling and release of the stored elastic energy. During these buckling and unbuckling processes sharp variations in the vertical velocity ($\pm$~600 m/s) are observed, which translate into high energy dissipation and hence enhanced friction (see SI sections 10 and 11). At higher normal loads (7.6 GPa), the magnitude of the vertical velocity bursts during buckling and unbuckling considerably diminishes (-60 - +30 m/s) and the contribution of the buckling process to the friction reduces (see SI Movie 3). Nevertheless, the overall frictional stress continues to grow in the high normal load regime, suggesting that another mechanism contributes to the energy dissipation.

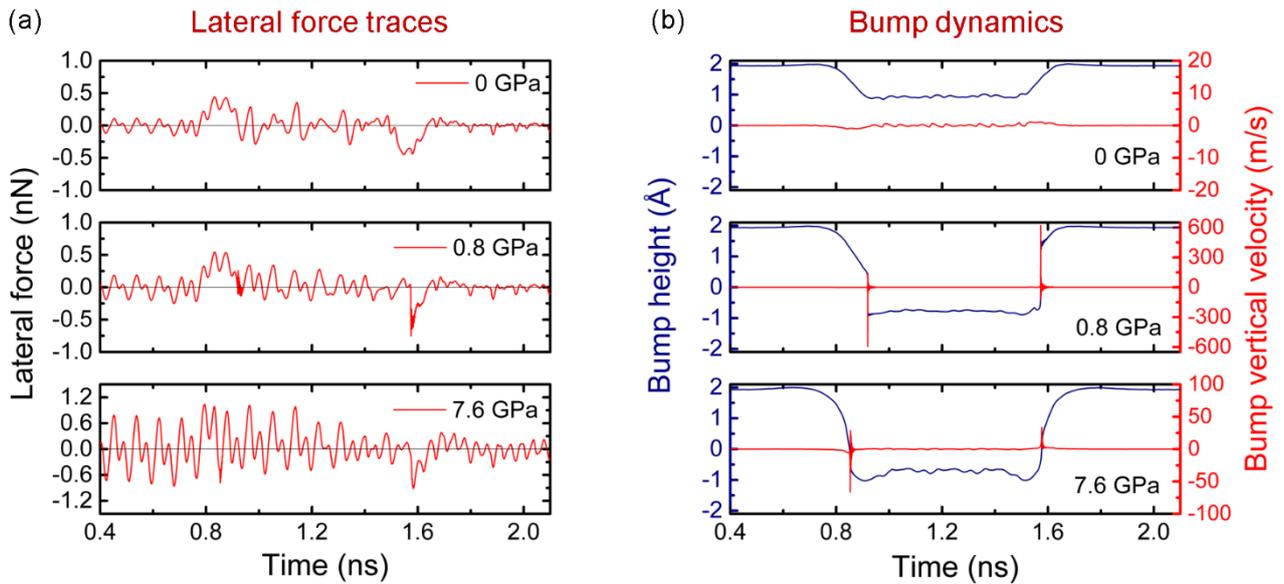

**Figure 4. Lateral force traces and bump dynamics of the $\theta = 4.7°$ GB in the forward sliding direction.** (a) Lateral force traces as a function of time at different normal loads. (b) The height (purple) and vertical velocity (red) of the bump as a function of time. Results for normal loads of 0, 0.8, and 7.6 GPa are presented in the top, middle, and low panels, respectively. The highest atom of the bump peak at the initial configuration is chosen for the calculations of bump height and bump vertical velocity at all simulation times. Additional results for lateral force traces and bump dynamics of the $\theta = 4.7°$, 13.9°, 27.8° GBs and the pristine graphene substrate can be found in SI section 12. See SI Movies 1-3 for the corresponding simulations of the $\theta = 4.7°$.

To elucidate this extra mechanism, we characterized the potential energy difference when the flake resides over the two grains (see SI section 13). To circumvent buckling effects we considered only the flat GB case ($\theta = 27.8°$). In Figure 5a, we plot the flake (black squares), interfacial (red circles), and substrate (blue triangles) components of this potential energy difference as a function of the normal load. It is clearly evident that the main contribution to the pressure dependence of the potential energy difference during the GB crossing process originates from the substrate. Notably, with increasing normal load the barrier for the GB crossing process grow linearly. To further elucidate this effect, we plot in Figure 5b the profile of the potential energy component associated with the substrate along the sliding path for various normal loads. Similar results for the other components can be found in SI section 13. We note that to reduce computational burden, these results were obtained by recording the corresponding potential energy contribution during a dynamical sliding simulation. We verified that this procedure faithfully represents results obtained via quasi-static calculations (see SI section 13). Notably, the pressure dependence of the substrate potential when the flake is located over Grain 2 is much more pronounced than for the case where the flake is located over Grain 1. This can be attributed to the different stacking modes of the two grains over the underlying pristine graphene surface resulting in different compressibility behaviors. Consequently, with increasing pressure, the potential barrier needed to be crossed by the flake, when moving from Grain 1 to Grain 2, increases. This, in turn, leads to a linear increase in the friction force with the normal load resulting in a differential friction coefficient of 0.001 (calculated as the derivative of the $\theta = 27.8°$ (blue triangles) curve in Figure 3c).

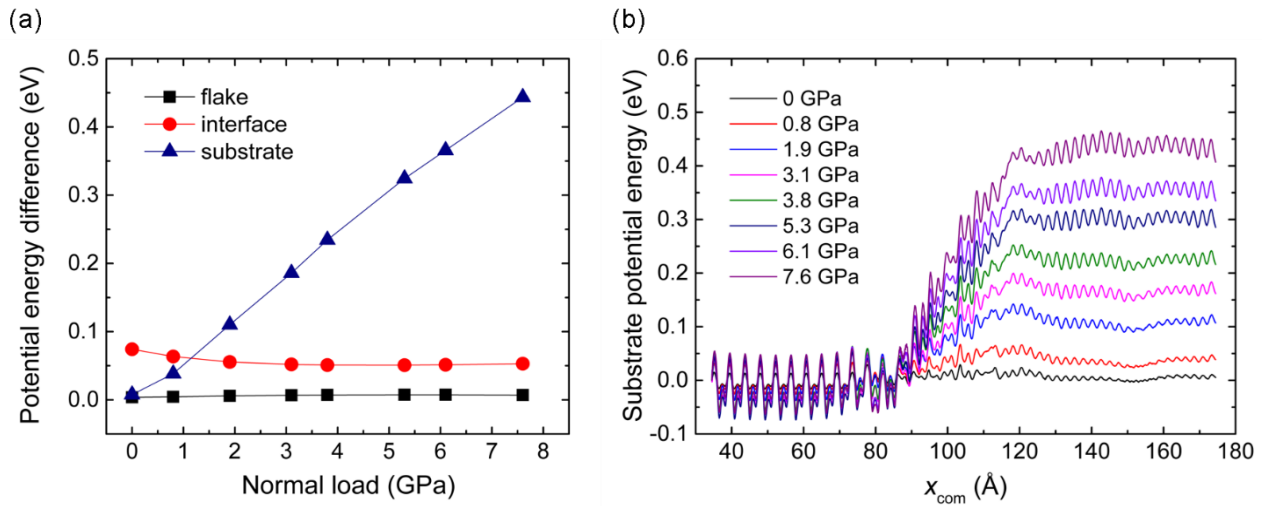

**Figure 5. Potential energy variations along the sliding path.** (a) The flake (black squares), interfacial (red circles), and substrate (blue triangles) energy contributions to the potential energy difference when the flake is removed from Grain 1 and placed on Grain 2. (b) The potential energy profiles of the substrate for $\theta = 27.8°$ as a function of the displacement of the center of mass of the top layer of the sliding flake at different normal loads. For comparison purposes, the substrate potential energy under each normal load when the flake is positioned deep inside Grain 1 is set to zero.

Importantly, this mechanism implies that when sliding over the GB in the opposite direction (from Grain 2 to Grain 1) the flake should experience an assistive force, whose magnitude increases with the normal load. A similar picture arises also for the corrugated GB systems discussed above, where bump buckling and grain compressibility effects are convolved. Therefore, in order to isolate the effect of the GB on the frictional properties of the system and to take into account that dynamic buckling can depend on the sliding direction we averaged the frictional stress obtained for the forward (see Figure 3c) and backward (Figure 3d) crossing process eliminating the grain compressibility contribution. This procedure corresponds well with typical friction experiments that measure multiple frictional loops.[6, 37, 38] Figure 3e presents the corresponding averaged frictional stresses as a function of normal load for the three GBs considered compared to a pristine substrate. The average friction experienced by the flake when crossing the flat GB is similar to that experienced when sliding over pristine graphene. The two corrugated GB systems show a completely different behavior from that of the flat GB case. At the low normal load regime, the GB bumps are merely slightly deformed under the slider in a reversible manner resulting in very low energy dissipation. As the normal load increases, the average frictional stress rises sharply due to the shear induced buckling discussed above. The corresponding dissipated energy has two contributions: (i) heat produced due to strong kinetic energy variations associated with the buckling process; and (ii) elastic energy stored in the buckled state. In case of unbuckling, the latter contribution can be partially recovered to produce an assistive force. Above an external normal load of ~2 GPa, the average frictional stress (after eliminating the potential gradient effect discussed above) gradually reduces with the external load approaching the frictional stress calculated over pristine graphene. This results from the fact that the vertical velocity bursts during the bump buckling and unbuckling process are suppressed by the external load and that buckling is found to be consistently reversible at this pressure regime. Overall, this behavior leads to the occurrence of negative average differential friction coefficients of the entire frictional loop[36] (see SI section 14 for the corresponding energy dissipation analysis).

The frictional mechanisms described above significantly differ from topographic mechanisms previously discussed in the literature. The latter associate friction and assistive forces with physical atomic-scale steps encountered by the slider along the surface.[39-41] On the contrary, our new findings suggest that, over corrugated GBs, friction may also originate from other sources including: (i) variations of compressibility along the surface, where different grains can store a different amount of elastic energy under the same normal load; (ii) heat produced during GB buckling and unbuckling events; and (iii) elastic energy storage in irreversible buckling processes. These unique mechanisms may lead to non-monotonic dependence of the average friction on the normal load and the occurrence of negative average differential friction coefficients.

**ACKNOWLEDGMENTS**

X.G. acknowledges the fellowship from the Sackler Center for Computational Molecular and Materials Science at Tel Aviv University. W.O. acknowledges the financial support from the Planning and Budgeting Committee fellowship program for outstanding postdoctoral researchers from China and India in Israeli Universities and the support from the National Natural Science Foundation of China (Nos. 11890673 and 11890674). O.H. is grateful for the generous financial support of the Israel Science Foundation under Grant no. 1586/17 and the Naomi Foundation for generous financial support via the 2017 Kadar Award. M.U. acknowledges the financial support of the Israel Science Foundation, Grant no. 1141/18 and the binational program of the National Science Foundation of China and Israel Science Foundation, Grant no. 3191/19.

## METHODS

### Polycrystalline Graphene Model Construction

The structures of polycrystalline graphene are generated by using a Voronoi tessellation method developed by Shekhawat,[28, 42] which creates physically realistic and low energy graphene GBs compared to the annealing method. Grain 1 is positioned such that its zigzag edge resides along the GB axis corresponding to an unrolled (0,1) nanotube of chiral angle $\theta_1 = 30°$ (see Figure 1a). Grain 2, is aligned such that its lattices corresponds to an unrolled ($n_1$, $n_2$) nanotube, where its chiral angle $\theta_2 = \arctan[(2n_1 + n_2)/\sqrt{3}n_2]$ is varied to generate GBs of misfit angles ($\theta = \theta_2 - \theta_1$) in the range of $0° - 60°$. The repeating unit cell along the grain boundary axis has translation vectors of $l_1 = \sqrt{3}a$ and $l_2 = a\sqrt{3(n_1^2 + n_2^2 + n_1 n_2)}$, for Grain 1 and Grain 2 respectively, where $a = 1.42039$ Å is the equilibrium carbon-carbon bond length. The corresponding lattice vector of the supercell along this direction is chosen to reduce the strain magnitude $|l_1 q - l_2 p|/|l_1 q + l_2 p|$ below $10^{-3}$, where $p$ and $q$ are the number of Grain 1 and Grain 2 duplicates in this direction within the supercell, respectively. To generate different GB configurations with the same misfit angle θ, the grain dimension perpendicular to GB is adjusted such that, the initial interface of the two triangle lattices for Grain 1 and Grain 2 is varied. After energy minimization of the interface and Voronoi tessellation, both flat and corrugated GB configurations are obtained.

### Grain Boundary Roughness Calculations

To evaluate the roughness of the various grain boundaries considered, the polycrystalline graphene surface was placed atop a pristine BLG substrate, which was taken to be periodic in both lateral directions with a supercell size of 6-20 nm along the GB axis and 10-20 nm in perpendicular direction. The zigzag edge of the substrate was placed along the GB axis. The polycrystalline graphene layer was taken to be periodic along the GB direction, whereas open boundary conditions were applied to this layer in the perpendicular direction to allow for lateral deformation. To that end, the length of the polycrystalline graphene in the direction perpendicular to the GB axis was taken to be ~1.7 nm shorter than the size of the box. The bottom layer of the BLG substrate was fixed rigidly during the calculation.

The geometry of the two top substrate layers was first optimized by using the FIRE algorithm[43] with a force convergence criterion of $10^{-3}$ eV/Å, followed by an annealing procedure, where the temperature was increased to 1000 K within 50 ps, maintained constant for 200 ps, and then cooled down to 0 K within 50 ps. Finally, the resulting structures were equilibrated for another 50 ps at 0 K. Here, the temperature of the system was controlled using Langevin dynamics with a damping coefficient of 1 ps$^{-1}$ applied to the middle substrate layer. To avoid getting stuck in local minima configurations, if a depressed bump geometry appeared after the above-mentioned procedure, another annealing round was performed until all the bumps protruded upwards. For the analysis of the bump density appearing in Figure 2b, an out-of-plane corrugation threshold of 0.5 Å was used to define the bump regions. The corresponding error bars in Figure 2 were evaluated by calculating at least three different structures for each misfit angle.

**Sliding Simulations**

For sliding simulations performed at zero temperature, the polycrystalline graphene substrate is generated and annealed in the same way as that in the roughness simulations. A TLG flake of lateral dimension 3.4×3.2 nm$^2$, with edge carbon atoms saturated by hydrogen atoms, is then placed atop Grain 1. To avoid a commensurate geometry, the orientation of the TLG flake is chosen to be rotated by 43.6° with respect to Grain 1 (exemplary geometry (xyz) files of the model systems are provided in the SI.). Normal load was applied by imposing a constant force to each carbon atom in the top layer of the flake (kept rigid, flat, and parallel to the fixed lower substrate layer throughout the simulations) in the direction perpendicular to the surface. The magnitude of the normal force was varied in the range 0-0.2 nN/atom, corresponding to pressures of 0-7.6 GPa, which are typical for tribological experiments and simulations.[5, 36, 44, 45]

The geometry of the entire system (apart from the fixed bottom substrate layer and the rigid top flake layer) is further optimized using the FIRE algorithm[43] with a force convergence criterion of 10$^{-3}$ eV/Å. Following the energy minimization, the flake was pulled by moving its top layer at a constant velocity of 5 m/s for 2.5 ns for GB misfit angles of $\theta = 4.7°$ and $\theta = 13.9°$, and 2.8 ns for $\theta = 27.8°$. Damped dynamics was applied to the middle substrate and flake layers with damping coefficients of 1 ps$^{-1}$ acting in all directions of motion of each damped atom. Furthermore, to reduce the effect of vertical oscillations of the entire flake an external damping force with 1 ps$^{-1}$ damping coefficient is also added to each atom of the top layer of the flake along the vertical direction, as suggested in Ref. 46. We verify that the results are insensitive to this choice of damping coefficient (see SI section 15). The friction force then is evaluated as minus the total instantaneous force exerted on the top layer of the flake.

To reduce computational burden the $\theta = 4.7°$ GB model was trimmed in the lateral directions and the dynamic simulations were performed with open boundary conditions. For $\theta = 13.9°$ and $\theta = 27.8°$ GBs, the polycrystalline graphene was kept periodic along the GB axis and open in the perpendicular (flake sliding) lateral direction.

For finite temperature simulations, the system was first equilibrated at 300 K using Langevin dynamics, then the flake was pulled along the same scan line as for the zero temperature simulations. Thermal averaging at each normal load value was done by performing three independent simulations with different initial conditions, generated consecutively during the equilibration simulations with 200 ps intervals.

**Quasi-Static Simulations**

Validation static and quasi-static simulations were performed to ensure that the energy profile estimations performed via dynamic simulations faithfully represent the quasi-static results (see SI section 13). The static calculations were performed by placing the TLG flake at different positions along the sliding path and relaxing the system under an external normal load while freezing the lateral motion of the top TLG flake layer. The quasi-static calculations were performed using the protocol proposed Bonelli *et al*.[47] We adopted the same system as that for the $\theta = 27.8°$ GB sliding simulations. The flake was initially positioned over Grain 1 (away from the GB region). At each step the top layer of the flake was rigidly displaced by 0.2 Å along the scan line towards Grain 2. The system was then relaxed (apart from the fixed bottom substrate layer and the rigid top flake layer) using the FIRE algorithm[43] with a force criterion of $2\times10^{-3}$ eV/Å. Due to extremely heavy computation cost, the quasi-static simulations were performed only at a typical normal load of 3.8 GPa scanning over two chosen segments in the scan line, which correspond to the regions of Grain 1 and the GB, respectively. The comparison of the energy profiles obtained by the static, quasi-static, and dynamic simulation shows negligible differences thus justifying the use of dynamical simulations to calculate the energy profile.

**Nudged Elastic Band Calculations**

To estimate the bump buckling energy barrier we performed nudge elastic band (NEB)[48-51] calculations (shown in SI section 9) between the unbuckled and buckled bump states extracted from the sliding simulations. To this end, the flake was first removed from the buckled and unbuckled snapshots and the system was allowed to relax using the FIRE algorithm[43] with a force criterion of $10^{-3}$ eV/Å. The resulting (local) minimum energy structures served as the initial and final structures for the NEB procedure and the buckling reaction coordinate was defined as per Ref. [52]. For the $\theta = 13.9°$ GB case, since there are intermediate states along the reaction path, the NEB calculations were performed in two steps, *i.e.* the reaction path between the first minimum and intermediate minimum, and the reaction path between the intermediate minimum and the final minimum.

**Additional Simulation Details**

All simulations were performed using the LAMMPS package.[53] The intralayer and interlayer interactions were modeled with the second-generation reactive empirical bond order (REBO) potential[54] and registry-dependent interlayer potential (ILP)[31-34] with refined parameters,[30] respectively. For roughness and sliding simulations, the time step used was 1 fs.

# Sliding Over Graphene Grain Boundaries: A Step Towards Macroscale Superlubricity

# Supporting Information

Xiang Gao, Wengen Ouyang, Oded Hod, and Michael Urbakh

Department of Physical Chemistry, School of Chemistry, The Raymond and Beverly Sackler Faculty of Exact Sciences and The Sackler Center for Computational Molecular and Materials Science, Tel Aviv University, Tel Aviv 6997801, Israel.

In this supporting information, we provide additional details on certain aspects of the study reported in the manuscript. The following issues are discussed:

1. Atomic Structures of Common Dislocations in Graphene
2. Effect of Out-of-Plane Deformation on Stress Field Relaxation
3. Atomic Structures and Topography of GBs with Respect to Misfit Angle
4. Calculation of GB Energy
5. Comparison of GB Energies of Corrugated and Flat GBs
6. Calculation of the Friction Force of GB
7. Effect of Scan Line on Friction
8. Temperature Effect on GB Friction
9. Nudged Elastic Band Calculations of the Elastic Energy
10. Calculation of Instantaneous Buckling Velocity
11. Kinetic Energy Change with Bump Buckling
12. Additional Sliding Simulation Results
13. Potential Energy Analysis
14. Energy Dissipation Analysis
15. Sensitivity Analysis of the Damping Rate Applied to the Top Layer of the Flake

# 1. Atomic Structures of Common Dislocations in Graphene

All the grain boundaries (GBs) generated in this study were constructed by three types of pentagon-heptagon (5|7) dislocations that can be described with the Burgers vector, $\vec{b}$, which is a topological invariant of the dislocation and a translational vector of graphene lattice, i.e. $\vec{b} = m\vec{a}_1 + n\vec{a}_2$, where $m$, $n$ are a pair of integers and $\vec{a}_1$, $\vec{a}_2$ are the graphene lattice vectors.[1] The pair of integers (m,n) is commonly used to denote the type of (5|7) dislocations, in analogy to the chiral indices of carbon nanotubes. In Figure S1, we show the atomic structures of three common types of (5|7) dislocations marked as (1,0), (1,1) and (2,0), which have been widely observed in experiments.[2,3] The dislocation can be viewed as the result of inserting a semi-infinite ribbon of width $|\vec{b}|$ into the otherwise perfect graphene lattice.[1] The (1,0) dislocation has the shortest Burgers vector, $|\vec{b}| = 2.46$ Å. The (0,1) dislocation, with a same length ($|\vec{b}| = 2.46$ Å.), differs from the (1,0) dislocation only in the direction of the Burgers vector (by 60°). Here, both (1,0) and (0,1) dislocations are referred as (1,0) type dislocations. A similar notation is used for (2,0) and (0,2) dislocations that both are referred as (2,0) type dislocations. The (1,1) dislocation has the second shortest Burgers vector $|\vec{b}| = 4.26$ Å and the (2,0) dislocation has a Burgers vector that is twice as long as that of the (1,0) dislocation.

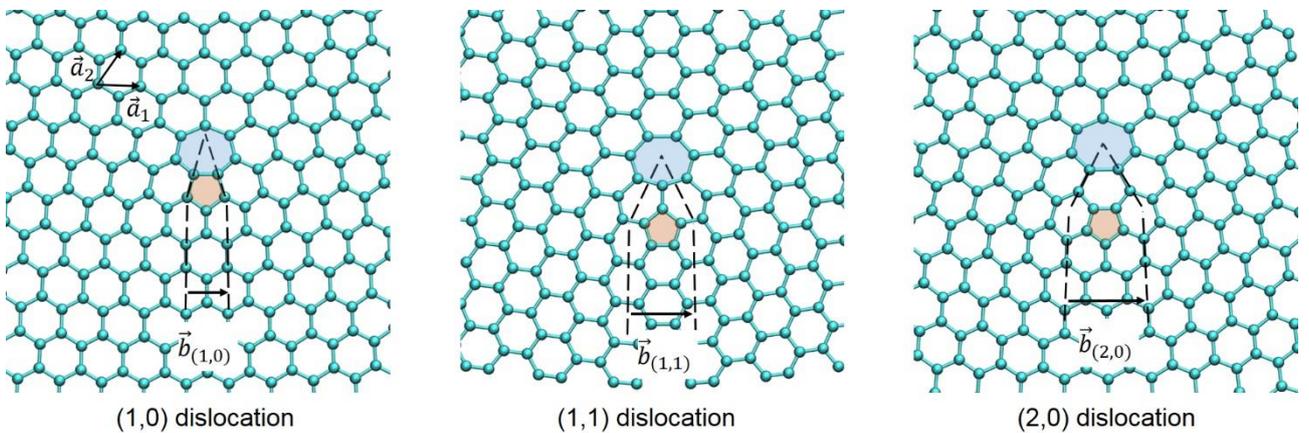

(1,0) dislocation      (1,1) dislocation      (2,0) dislocation

**Figure S1.** Atomic structures of (1,0), (1,1) and (2,0) dislocations. The pentagon and heptagon defects are filled with light orange and light blue, respectively. The dashed lines outline the inserted semi-infinite graphene ribbons. For clarity of the presentation, the Burgers vectors of the three illustrations are aligned in the same direction.

## 2. Effect of Out-of-Plane Deformation on Stress Field Relaxation

To explain why some GB structures result in corrugated seam lines we performed geometry optimization with and without out-of-plane relaxation for the $\theta = 9.7°$ GB (Figure 1a of the main text) and studied the corresponding in-plane stress field. Figure S2 shows the color maps of the distribution of C-C bond length and in-plane stress with and without out-of-plane deformation. For the case without out-of-plane deformation (left column), the atoms in the polycrystalline graphene are relaxed only in lateral directions, whereas for the case with out-of-plane deformation (right column), the atoms in the polycrystalline graphene are allowed to relax in all three directions. The relaxation protocol is the same as that described in the Method section of the main text.

It is seen that, without out-of-plane deformation, the dislocations in the GB lead to strongly compressed and stretched regions alternately distributed along the GB (Figure S2a, e, and g). The compressed regions and the stretched regions correspond to the pentagons and heptagons of the dislocations (see Section 1 above), respectively. With out-of-plane deformation, the compressed pentagons protrude upward forming surface bumps (Figure S2c,d). Correspondingly, the compressed C-C bond length and the compressive in-plane stress at the pentagons are significantly relaxed (Figure S2b, f, and h).

The dependence of the bump height on the GB misfit angle $\theta$ can be rationalized considering the alternating compressed and stretched regions in the vicinity of the GB seam. The stress field of a dislocation consists of compression on the pentagon side and tension on the heptagon side, which is relieved by the stress fields of its neighboring dislocations. As $\theta$ increases, the distance between neighboring dislocations decreases, therefore promoting the stress tension cancellation. This effect is enhanced when the two dislocations are directly connected through sharing a C-C bond, hexagon ring, or a pair of mutual edge atoms. Thus, the bump height decreases with increasing $\theta$, and the bumps can be even annihilated when directly connected to their neighboring dislocations.

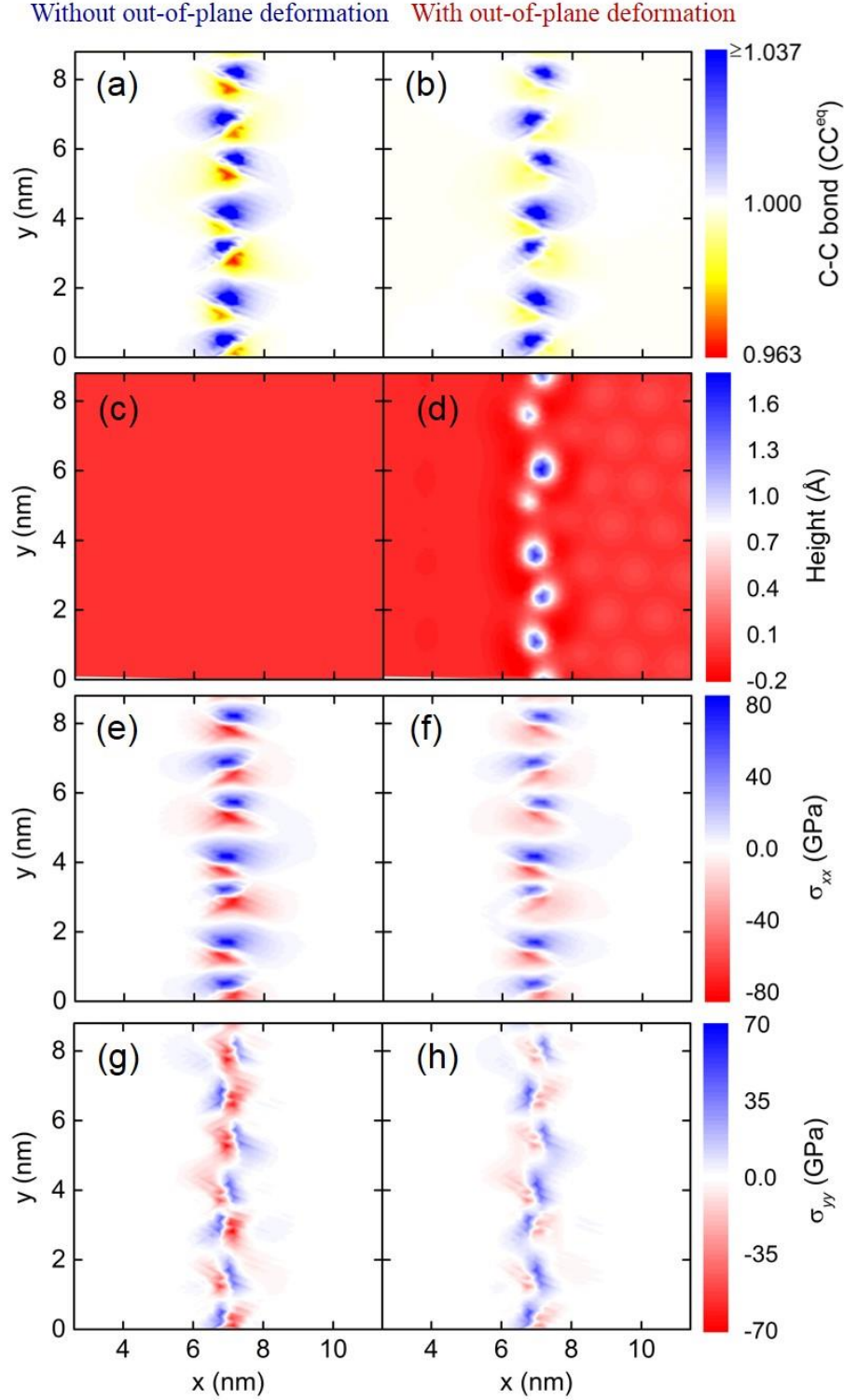

**Figure S2.** Effect of out-of-plane deformation on the stress field relaxation for a $\theta = 9.7°$ GB. (a)–(b) the bond length distribution; (c)-(d) height distribution, (e)-(f) $x$-axis stress distribution, and (g)-(h) $y$-axis stress distribution. Panels in the left column are relaxed without allowing for out-of-plane deformation whereas panels in the right column are relaxed allowing for out-of-plane deformation. The C-C bond length is normalized by the equilibrium carbon-carbon distance in an isolated pristine graphene surface, $a_{CC} = 1.42039$ Å. Using the procedure described in Ref. 4, the stresses are calculated from the global stress tensor obtained from LAMMPS divided by the atomic volume of carbon atoms. The latter is evaluated with $3\sqrt{3}a_{CC}h/4$, where $h = 3.35$ Å is the equilibrium interlayer distance.

## 3. Atomic Structures and Topography of GBs with Respect to Misfit Angle

Figure S3 reports a series of atomic structures and topography of Corrugated-I GBs (corrugated with only (1,0) type dislocations) with misfit angles $\theta$ = 3.3°, 6.6°, 12.5°, 18.1° and 24.3°, respectively. As $\theta$ increases, the density of dislocations increases while the height of bumps drops due to stress field cancellation discussed in Section 2. For $\theta < 12.5°$, the bump density $D$ increases linearly with $\theta$ as each dislocation generates a surface bump. While for $\theta \geq 12.5°$, the dislocations start to connect directly with each other through sharing a C-C bond, hexagon carbon ring, or a pair of mutual edge atoms, which significantly reduces the local stress field and annihilates bumps. Thus, the bump density decreases with increasing $\theta$ in this angle range. Flat GBs (with a typical surface corrugation lower than 0.2 Å as shown in Figure 2a in the main text) are observed at misfit angles of $\theta$ = 21.8°, 25.7°, 27.8° and 30°. Flat-I GBs (flat with only (1,0) type dislocations) are observed at all the four misfit angles, while Flat-II GBs (flat with (1,0) type dislocations mixed with (1,1) or (2,0) type dislocations) are only found at $\theta$ = 27.8° (with (1,1) type dislocation). For the flat $\theta$ = 21.8° GB, the dislocations are connected with each other by sharing a C-C bond, a hexagon carbon ring, or a pair of mutual edge atoms. For the flat $\theta$ = 25.7° and 27.8° GBs, the dislocations are arranged by sharing a C-C bond or a pair of mutual edge atoms. While for $\theta$ = 30°, the dislocations form a fully edge-sharing dislocation chain.

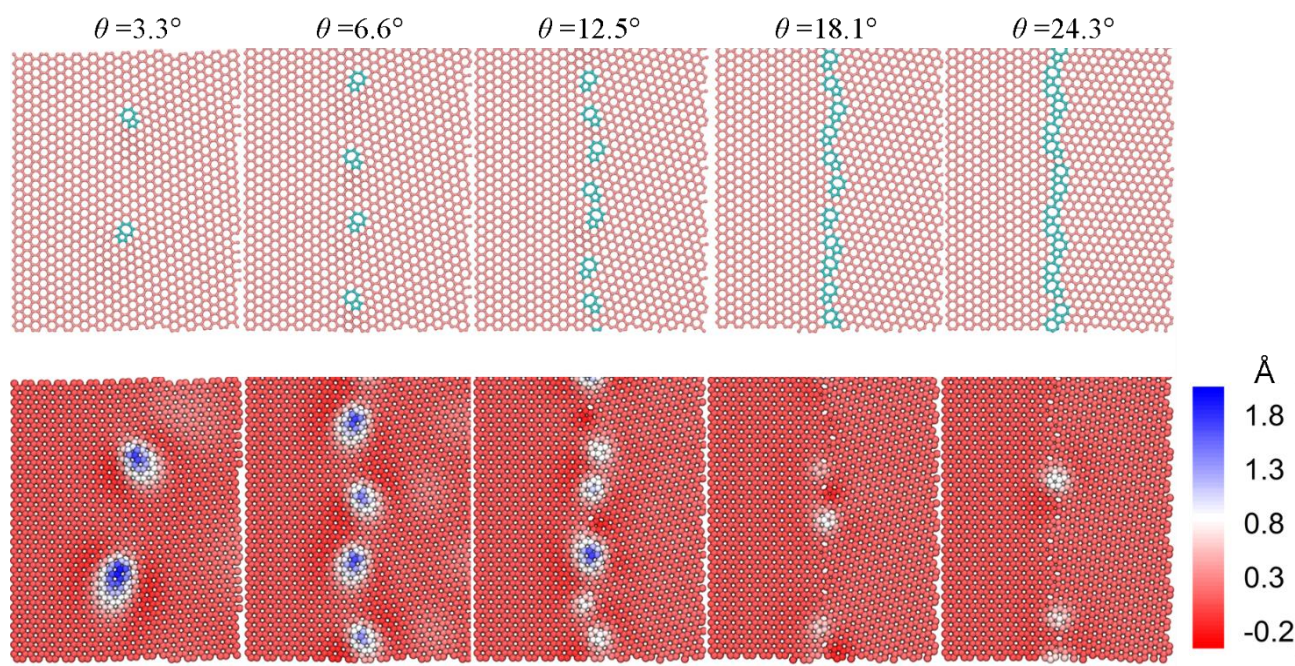

**Figure S3.** Atomic structures and topography for corrugated GBs (Corrugated-I) at misfit angles of $\theta$ = 3.3°, 6.6°, 12.5°, 18.1° and 24.3°, respectively. Panels in the top row show the atomic structures of GBs on a graphite substrate after annealing. Panels in bottom row report the corresponding topography (height distribution) of the GBs in the top row. The color of atoms represents the atomic height with respect to the average height of the two grains.

The appearance of flat GB at misfit angles of $\theta$ = 21.8°, 25.7°, 27.8° and 30° is consistent with previous density functional theory calculations of free standing polycrystalline graphene at GB misfit angles of 21.8°, 32.2° and 38.2°,[1] as well as STM measurement of graphene supported on a SiC substrate with misfit angle in the range of 25° < $\theta$ < 40°.[5]

In addition, for the four misfit angles where flat GBs appear, when the stress field is not sufficiently cancelled along the GB seamline corrugated GBs (Corrugated-I and Corrugated-II) can also form. Figure S4, shows four examples of such corrugated GBs.

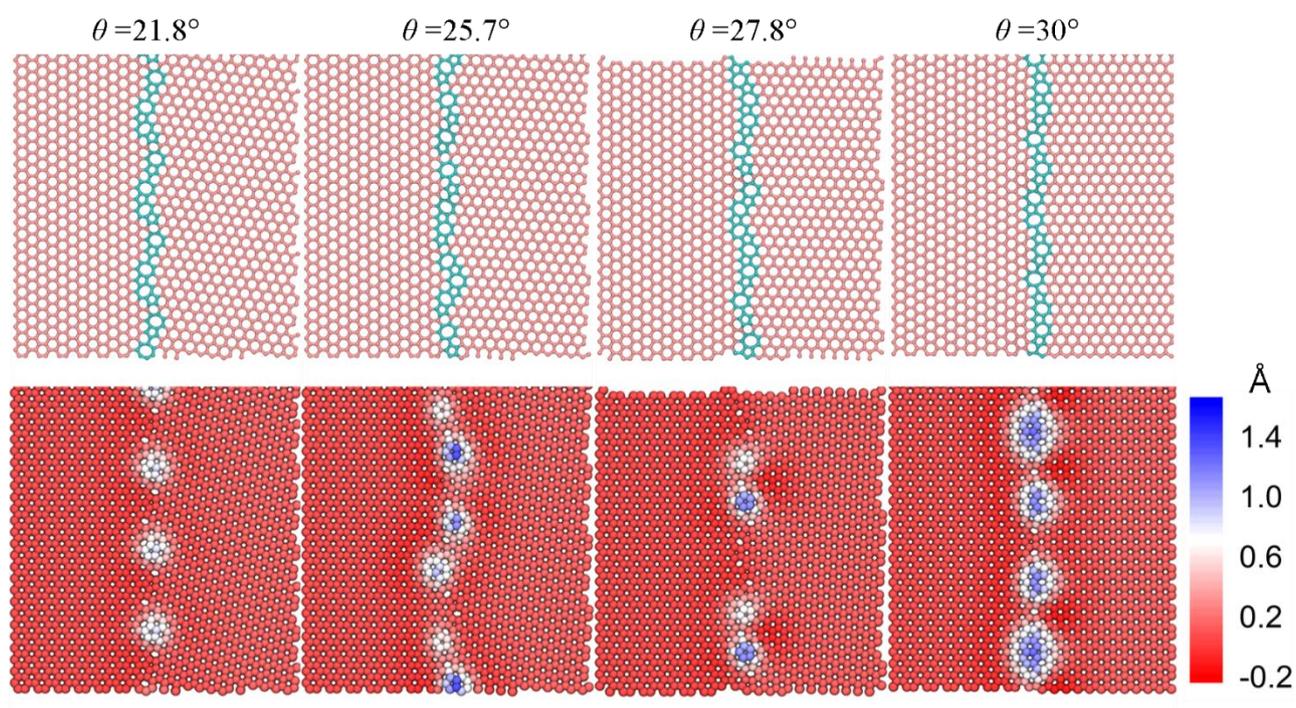

**Figure S4.** Atomic structures and topography of corrugated GBs (Corrugated-I and Corrugated-II) at misfit angles of $\theta$ = 21.8°, 25.7°, 27.8° and 30°, respectively. Panels in top row show the atomic structures for GBs on a graphite substrate after annealing. Panels in the bottom row present the corresponding topography of the GBs in the top row. The color represent the atomic height with respect to the average height of the two grains.

## 4. Calculation of GB Energy

To evaluate the GB energy per unit length, $\gamma$, we adopt the approach used in Ref. 2, where $\gamma = (E_{\text{total}} - n_{\text{atoms}} \times \bar{E}_{\text{bulk}})/l_{\text{GB}}$. Here $E_{\text{total}}$ is the total potential energy of the atoms in the polycrystalline graphene layer, $n_{\text{atoms}}$ is the total number of atoms in the polycrystalline graphene layer, $\bar{E}_{\text{bulk}} \approx 7.42$ eV (Figure S5) is the average potential energy per atom in the bulk regions of Grain 1 and Grain 2, and $l_{\text{GB}}$ is the length of the GB. The calculated GB energy per unit length $\gamma = 0 - 0.5$ eV/Å is comparable to that obtained for buckled GBs in free-standing polycrystalline graphene $(\gamma = 0 - 0.4$ eV/Å$)$.[1]

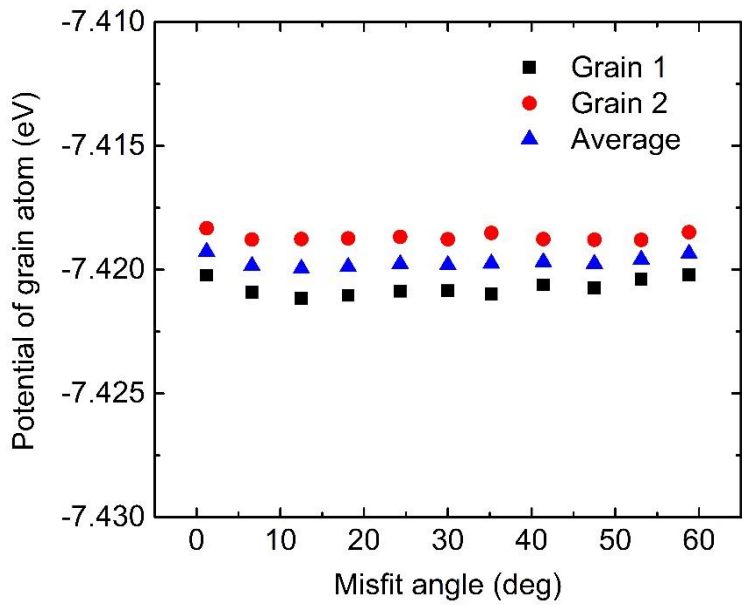

**Figure S5.** The potential energy per atom in the bulk regions of the two grains and their average as a function of misfit angle.

## 5. Comparison of GB Energies of Corrugated and Flat GBs

In Figure 2 of the main text, the energies of the Corrugated-I and Flat-I GBs are roughly identical. The reason is that, after geometry relaxation, the GB energy is mainly determined by the type (e.g. (1,0), (1,1), or (2,0) type) and density of the dislocations. For Corrugated-I and Flat-I GBs with same misfit angle, the type and density of dislocations are similar, as shown in Figure S6. Prior to geometry relaxation, Flat-I type GBs show lower energy than their Corrugated-I counterparts, because the latter exhibits larger in-plane stress. Following relaxation, the in-plane stress is partially released and the energy of the Corrugated-I GBs decreases, approaching the value of Flat-I GBs. A similar picture arises for the Corrugated-II and Flat-II GBs. To exemplify this effect, we calculated the energy of the $\theta = 21.8°$ Corrugated-I GB. For this system, the GB energy drops from 0.51 eV/Å prior to geometry relaxation to 0.46 eV/Å following relaxation. The latter is close to the GB energy of the Flat-I GB with the same misfit angle.

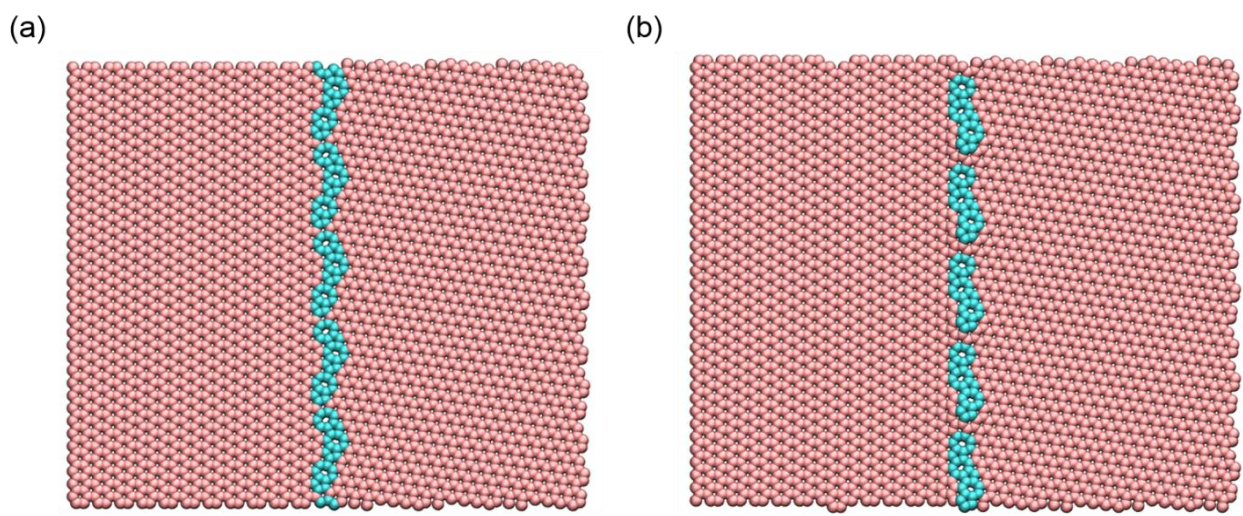

**Figure S6.** Atomic structures of (a) a Flat-I GB and (b) a Corrugated-I GB with a misfit angle $\theta = 21.8°$. The cyan and pink spheres represent the pentagon-heptagon dislocation atoms and hexagonal carbon atoms, respectively.

## 6. Calculation of the Friction Force of GB

The friction force is calculated by averaging the lateral force experienced by the top rigid layer of the tri-layer graphene (TLG) flake in a predefined region around the GB seam. This region is defined as the region where the force trace deviates from its typical behavior deep inside the contacting grains. For example, the lateral force trace for the $\theta = 4.7°$ GB under zero normal load is shown in Figure S7. The region between the dash lines is defined as the GB region of width ~5.9 nm. The same region is used for all the friction calculations under different normal loads for both sliding directions. We note that the qualitative frictional behavior presented in the main text is not sensitive the exact choice of GB region. A similar procedure is used for all other GBs considered in this study.

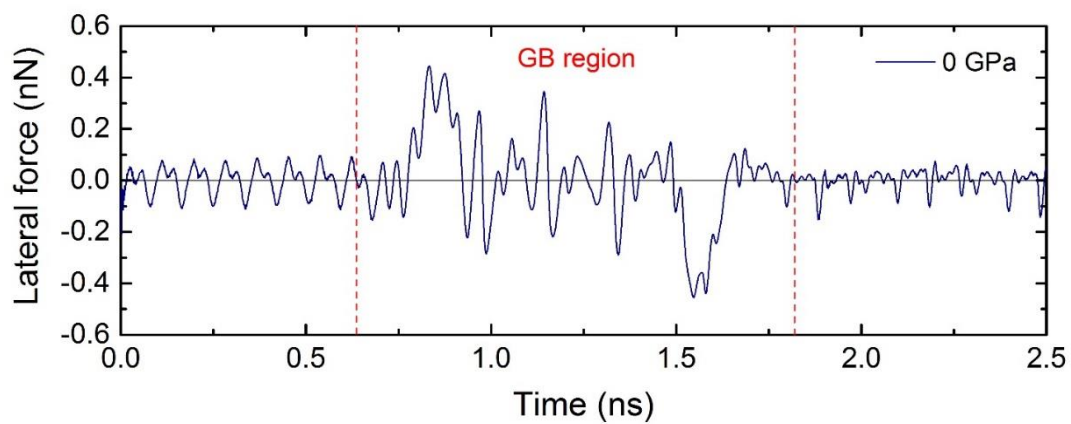

**Figure S7.** GB region definition. The lateral force trace as a function of simulation time at zero normal load is presented for the $\theta = 4.7°$ GB. The red dash lines mark the defined boundaries of the GB region.

## 7. Effect of Scan Line on Friction

The scanline for the $\theta = 4.7°$ GB was deliberately chosen so that the flake would slide over a single GB bump avoiding interaction with adjacent bumps. This allowed us to assess the effects of a single GB defect on the frictional properties of the system. To examine the effect of scanning off-center with respect to a GB defect, we chose a scanline shifted downwards by 5 Å along the GB axis with respect to the original one (see Figure S8a,b). Along this scan line the flake is still strongly influenced by the original bump but has some overlap with the adjacent bump. For comparison purposes, we performed simulations at representative normal loads of 0, 0.8, and 2.3 GPa. Similar to the results obtained along the original scan line, we observe bump buckling at both 0.8 and 2.3 GPa but not at 0 GPa. The comparison of frictional stresses between the original scan line and shifted scan line are shown in Figure 8c, d and e, for forward sliding, backward sliding, and their average, respectively. It can clearly be seen that there is good qualitative agreement between the original and the new results. Therefore, we conclude that the scan line has minor effect on the general qualitative frictional behavior of the system and on the underlying mechanisms.

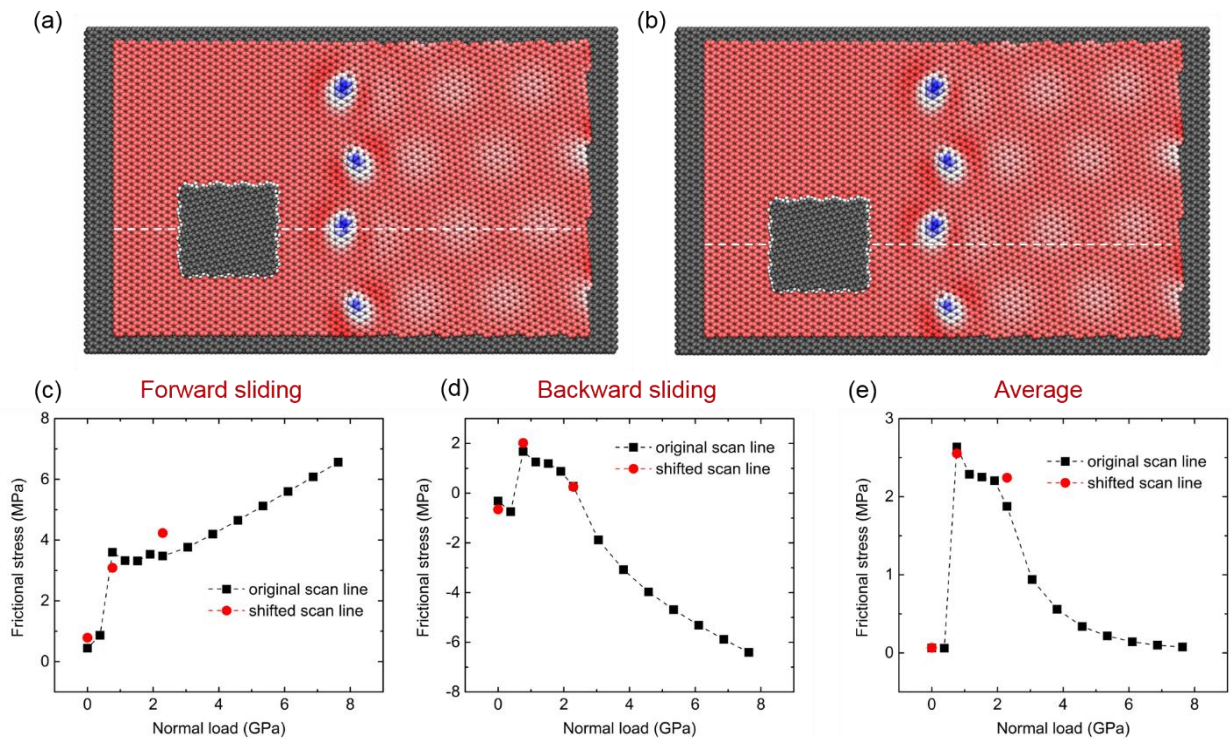

**Figure S8.** Effect of the slider scan-line on the friction. (a) Model system of misfit angle $\theta = 4.7°$ with the scan line (dashed white) used in the main text. (b) Same as (a) with a scan line shifted downward by 5 Å. (c)-(e) Comparison of the frictional stresses obtained using the original scan line in (a) and shifted scan line in (b), as a function of normal load in the forward and backward sliding directions and their average, respectively.

## 8. Temperature Effect on GB Friction

To demonstrate the temperature effect on the GB friction, we performed forward sliding simulations over the $\theta = 4.7°$ GB at 300 K. The comparison of the shear stress at 0 K and 300 K is shown in Figure S9. Generally, the shear stress at 300 K behaves similar to the 0 K case. At low normal loads (< 1.5 GPa), the shear stress increases rapidly with normal load and roughly follows the results of 0 K. This effect is dominated by bump buckling events at both temperatures considered. At the high normal load regime (> 4 GPa), the shear stress is slightly higher than that calculated at 0 K, showing a similar increase with the normal load. We note that at 300 K the critical normal load for bump buckling is reduced to nearly zero and the frequency of buckling events is enhanced due to thermal fluctuation.

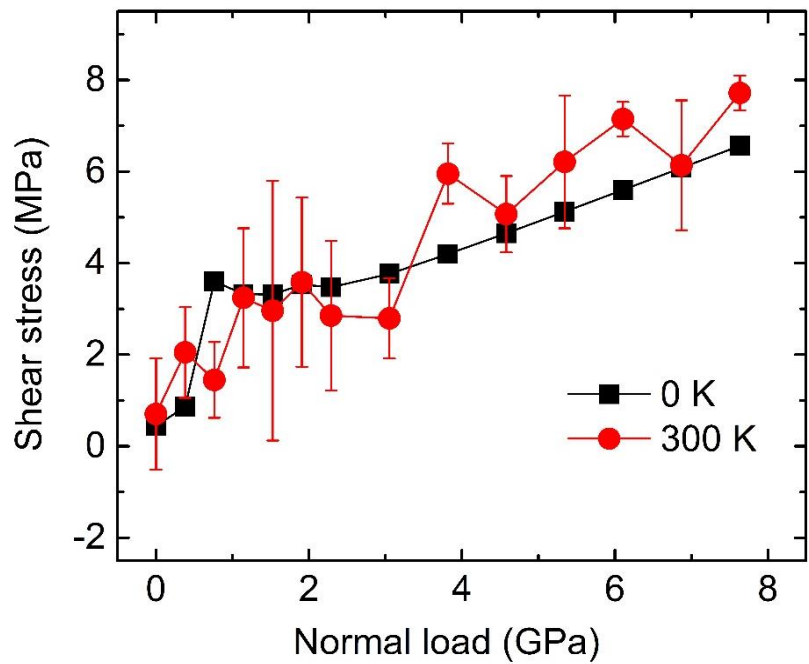

**Figure S9.** Shear stress in forward sliding over the $\theta = 4.7°$ GB at 0 K and 300 K as a function of normal load. The error bars for 300 K are obtained by averaging three individual simulations with different initial conditions.

## 9. Nudged Elastic Band Calculations of the Elastic Energy

To evaluate the amount of elastic energy that can be stored in the GB due to buckling and the corresponding barriers we performed nudged elastic band (NEB) calculations[6-10] along individual bump buckling reaction coordinate for the corrugated GBs considered in the absence of the sliding flake (see details in the Method Section). Figure S10a presents the buckling potential energy profile for $\theta = 4.7°$ GB, where the insets show the configuration of the (a) unbuckled, (b) transition, and (c) buckled states. The energy barrier for downward buckling is found to be ~0.9 eV explaining the abovementioned normal load threshold required for buckling to occur. The considerably lower (~0.1 eV) unbuckling barrier explains why buckling is found to be reversible in many of our dynamic simulations. We note however, that dynamical effects in the presence of the sliding flake can results in irreversible buckling (see SI Movie 4), leading to the loss of energy towards elastic degrees of freedom, therefore enhancing friction.

For the $\theta = 13.9°$ high bump density GB case a somewhat more involved picture arises. Here, the sliding area contains more than one bump and hence several reaction paths can be envisioned exhibiting different bump buckling sequences. Two such sequences are demonstrated in Figure S10b,c where the reaction paths, barriers, and corresponding geometries depend on which of the two bumps residing in the sliding path buckles first.

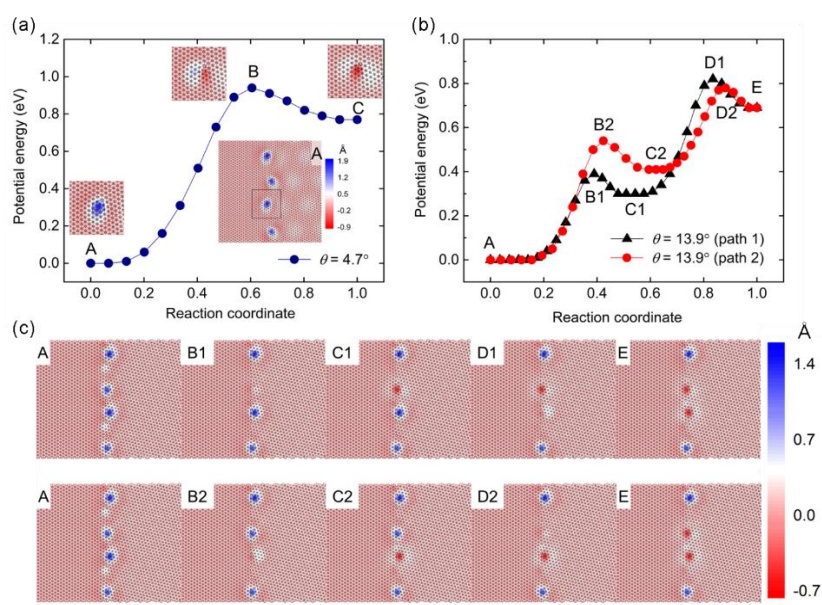

**Figure S10.** NEB calculations. (a) Potential energy of the system along the reaction coordinate for the $\theta = 4.7°$ GB buckling without the TLG flake. The black square in the inset marks the bump which resides along the sliding path. The zoomed-in snapshots are displayed near each minimum and the transition state. (b) Potential energy of the system along the reaction coordinate for the $\theta = 13.9°$ GB. (c) The snapshots for each minimum and transition state appearing in (b). The lowest energy state along the buckling reaction coordinates for both systems are shifted to zero for clarity of the presentation. The color scales denote the atomic height with respect to the average height of the two grains.

## 10. Calculation of Instantaneous Buckling Velocity

The instantaneous bump velocity in the vertical direction $v_t$ at time $t$ is calculated as $v_t = \frac{1}{\Delta t}(z_{t+\Delta t/2} - z_{t-\Delta t/2})$, where $\Delta t$ is the time interval, and $z_{t+\Delta t/2}$ and $z_{t-\Delta t/2}$ are the $z$ coordinates of the peak atom of the bump (chosen at the initial configuration) at time $t + \Delta t/2$ and $t - \Delta t/2$, respectively. The calculated velocity profiles with different time intervals for a buckling event are shown in Figure S11. In the region away from the buckling position, a time interval of 1 ps is used (not to be confused with the simulation time-step of 1 fs), which shows minor effect on the velocity calculation. While in the buckling region, the fast motion associated with the buckling process is not well captured using the same time interval, which significantly underestimates the instantaneous buckling velocity. To provide sufficient numerical accuracy of the bump buckling simulation, we used a 100 fs time interval to calculate the instantaneous velocity in the vicinity of the buckling region. Then, the full velocity profile is built by combining the parts in the buckling region and away from the buckling region.

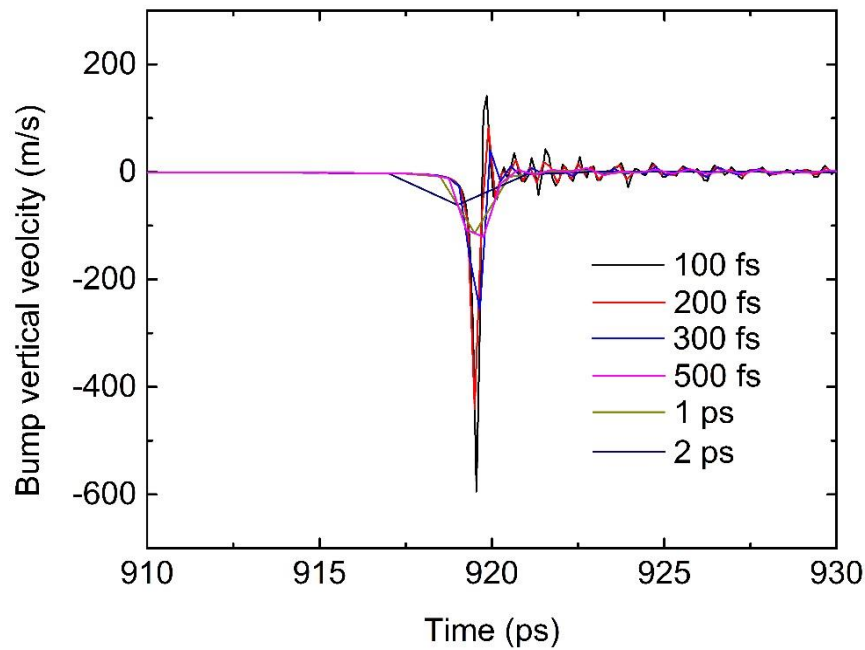

**Figure S11.** Instantaneous velocity profiles for a buckling event calculated using different time intervals for the $\theta = 4.7°$ GB at a normal load of 0.8 GPa.

## 11. Kinetic Energy Change with Bump Buckling

Figure S12a shows the typical kinetic energy change of the system during a buckling event. The total kinetic energy jumps up suddenly at certain point and then gradually drops with some fluctuations. The sudden jump of the total energy results from the instantaneous buckling velocity (as high as ~600 m/s under our simulation conditions), as shown in Figure 4b in the main text. To further illustrate this, we calculate separately the kinetic energy components of the flake, in the bump region and non-bump region, as illustrated in Figure S12b. The bump region is defined as a circular region centered at the peak atom of the bump (chosen at the initial state) with a radius of 1 nm (sufficiently large to cover most of the bump region based on the height profile while avoiding overlap with other bumps, as shown in Figure S12c). The atoms in the same circular region in the layer below the polycrystalline graphene are also included in the bump region. The rest of the substrate is defined as the "non-bump" region. The assignment of atoms to the two regions is made according to the initial state and remains unchanged during the simulation. As shown in Figure S12b, a kinetic energy spike is first generated in the bump region, and then it transfers to the flake and non-bump region. The dissipation of this kinetic energy pulse lasts for ~5 ps. This clearly supports the argument that the pulse in total kinetic energy of the system is caused by bump buckling.

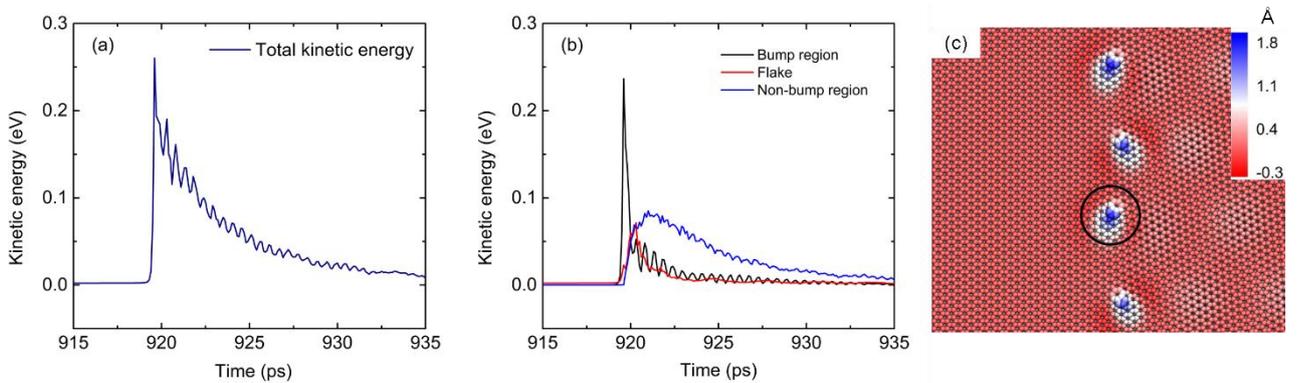

**Figure S12.** Kinetic energy variations during a buckling event for the $\theta = 4.7°$ GB under a normal load of 0.8 GPa. (a) Total kinetic energy of the system. (b) Kinetic energy of the bump region, the flake, and "non-bump" region. (c) Definition of the bump region, which is outlined with the black circle. The colors of the atoms represent the vertical height of the atoms above the average height of the two grains.

## 12. Additional Sliding Simulation Results

### 12.1. Backward Sliding over the corrugated $\theta = 4.7°$ GB

Following the forward sliding simulations, where the TLG flake is initially positioned over Grain 1 and pulled towards Grain 2, we performed backward sliding simulations starting from the endpoint located over Grain 2 and pulling the TLG flake towards Grain 1 along the same scanline and at the same pulling velocity of 5 m/s. The lateral force traces (left column) and bump dynamics (right column) at different normal loads during the backward sliding over the $\theta = 4.7°$ GB are shown in Figure S13. In the backward sliding simulation, the bump was found to buckle at 0.8 GPa (see panels c and d of Figure S13, and SI Movie 4), similar to the forward sliding case. However, in contrast to the forward sliding dynamics, for loads below 3 GPa the bump does not unbuckle but rather remains depressed even after the flake leaves the GB region (see panel d of Figure S13, and SI Movie 4). This is clearly reflected in the force trace at 0.8 GPa normal load in panel c of Figure S13, where the flake experiences a resistive force when first encountering the GB (0.86-0.96 ns in the diagram) followed by force oscillations corresponding to the downward bump buckling (at simulation time of 0.96 ns). Notably, the negative lateral force observed in the forward sliding case (see Figure 4a of the main text) at this pressure range due to bump unbuckling is lacking. At a normal load of 3 GPa the bump unbuckles after the TLG flake leaves the GB region resulting in sharp force spikes at simulation time of 1.67 ns, followed by force oscillations (see panels e and f of Figure S13). A similar behavior is found at higher normal loads, however since the unbuckling bump vertical velocity in this case is reduced, the sharp spikes observed at 3 GPa are replaced by smoother oscillations.

The different behavior between the forward and backward sliding dynamics can be attributed to the asymmetry of the bump structure with respect to the GB axis, where the pentagonal dislocation rings incline towards Grain 1 (see Figure S14). We find that, once buckled, the bump is more likely to unbuckle when the flake leaves the GB from the heptagon side (*i.e.* in forward sliding direction) than when it leaves from the pentagon side (backward sliding direction).

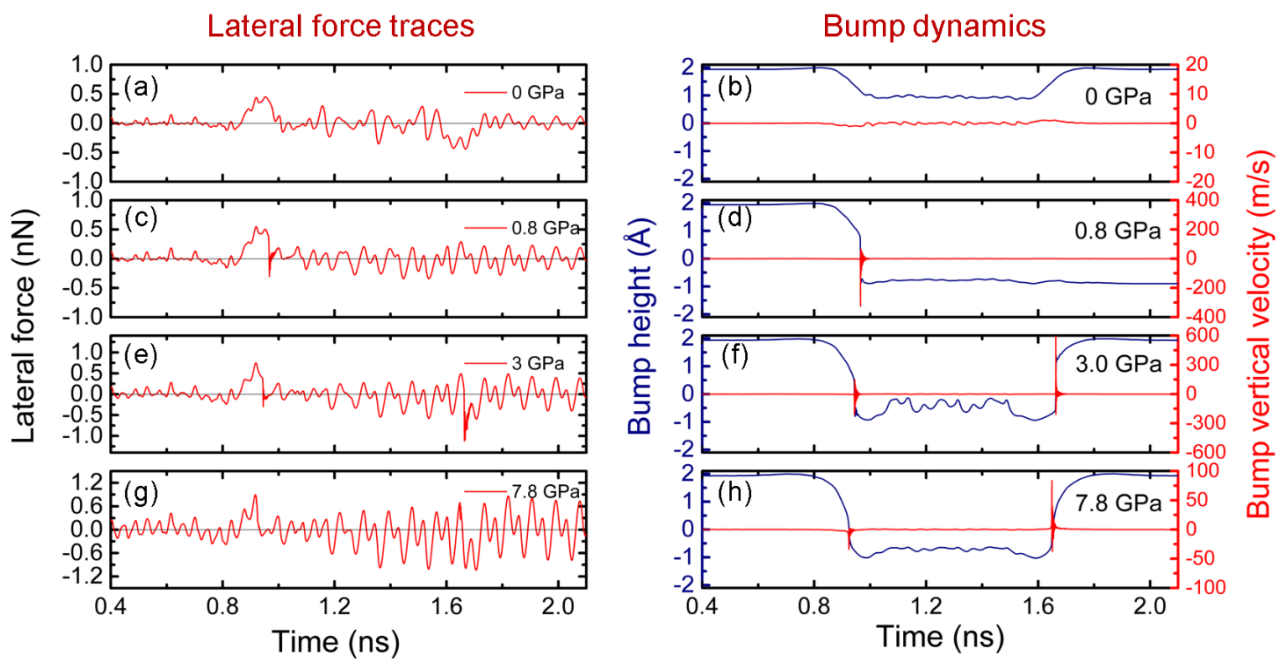

**Figure S13.** Lateral force traces and bump dynamics in the backward sliding direction for the $\theta = 4.7°$ GB. Panels (a), (c), (e), and (g) present the lateral force traces as a function of time for normal loads of 0, 0.8, 3.0, and 7.8 GPa, respectively. Panels (b), (d), (f), and (h) show the corresponding bump heights and bump vertical velocities variation profiles. Positive values indicate a resistive force.

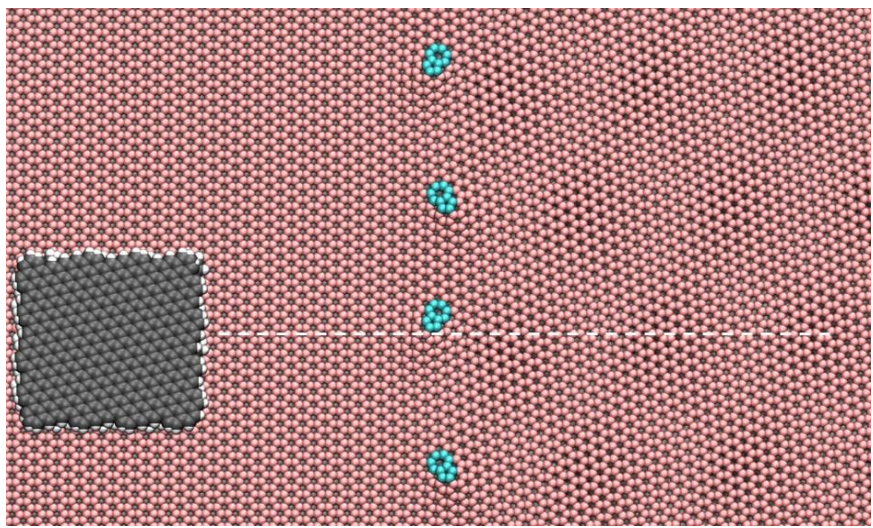

**Figure S14.** Illustration showing the orientation of the pentagon-heptagon defects along a $\theta = 4.7°$ GB with respect to the sliding path of the TLG flake. The grey and while spheres represent the carbon and hydrogen atoms of the slider, respectively. The pink and cyan spheres depict the hexagon and dislocation polycrystalline substrate atoms, respectively. The white dashed line depicts the scan line, where the sliding direction is from left to right.

## 12.2. Forward and Backward Sliding over the corrugated $\theta = 13.9°$ GB

The sliding simulation setup for the $\theta = 13.9°$ GB is shown in Figure S15. For this system, the TLG flake slides simultaneously over two bumps of heights 1.5 Å (bump 1) and 1.4 Å (bump 2). The lateral force traces at different normal loads in the forward sliding direction are shown in the left column of Figure S16. The corresponding height dynamics of the two bumps is shown in the right column of the figure. Under zero normal load neither of the bumps buckle during the TLG flake crossing of the GB (see panels a and b of Figure S16, and SI Movie 5). A resistive force resulting from the downward bump depression at the simulation time of 0.92-1.12 ns (or sliding distances of 4.6-5.6 nm) followed by enhanced force oscillations and then a negative lateral force due to bump recovery at a simulation time of 1.64-1.8 ns (sliding distances of 8.2-9.0 nm), can be clearly seen. At the normal load of 0.4 GPa (see panels c and d of Figure S16, and SI Movie 6), both bumps buckle downward and remain buckled after the TLG flake leaves the GB region. Consequently, the negative lateral force associated with bump unbuckling disappears. The buckling of the bumps leads to the rapid increase of friction observed in the low normal load regime of Figure 3c in the main text. As the normal load is increased to 2.3 GPa, bump 2 unbuckles when the TLG flake leaves the GB region (resulting in the reappearance of a negative lateral force), whereas bump 1 remains buckled (see panels e and f of Figure S16, and SM Movie 7). For normal loads greater than 4.6 GPa, both bumps unbuckle when the TLG flake leaves the GB region (see panels g and h of Figure S16, and SI Movie 8).

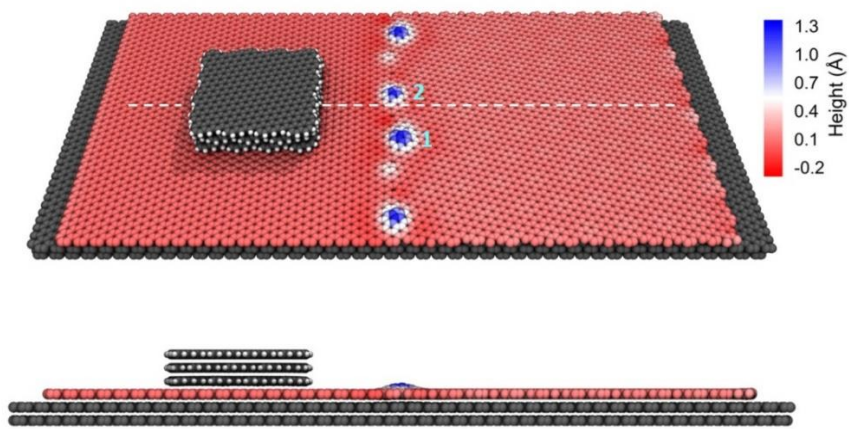

**Figure S15.** Sliding simulation setup for the $\theta = 13.9°$ GB. Both perspective (top panel) and front (bottom panel) views are provided. The two bumps in the sliding path, *i.e.* bump 1 and bump 2, are explicitly denoted in the top panel. The grey and white spheres represent carbon and hydrogen atoms, respectively. Coloring of the polycrystalline atoms represents their height with respect to the average height of the two grains (see color scale). The white dashed line depicts the scan line, where the sliding direction is from left to right.

The lateral force traces at different normal loads in the backward sliding direction over the $\theta = 13.9°$ GB are shown in the left column of Figure S17. A qualitatively similar picture to that found for the forward sliding direction is obtained, where bump 1 and 2 buckle (and remain buckled even after the TLG flake leaves the GB region) at a normal load of 0.4 GPa. Notably, the bumps unbuckle at lower normal loads (1.9 GPa and 2.3 GPa for bump 1 and bump 2, respectively) than those found for the forward sliding. This reflects the asymmetry of the bumps with respect to the GB axis, similar to the case of the $\theta = 4.7°$ GB. See SI Movie 9 for the case of 1.9 GPa normal load.

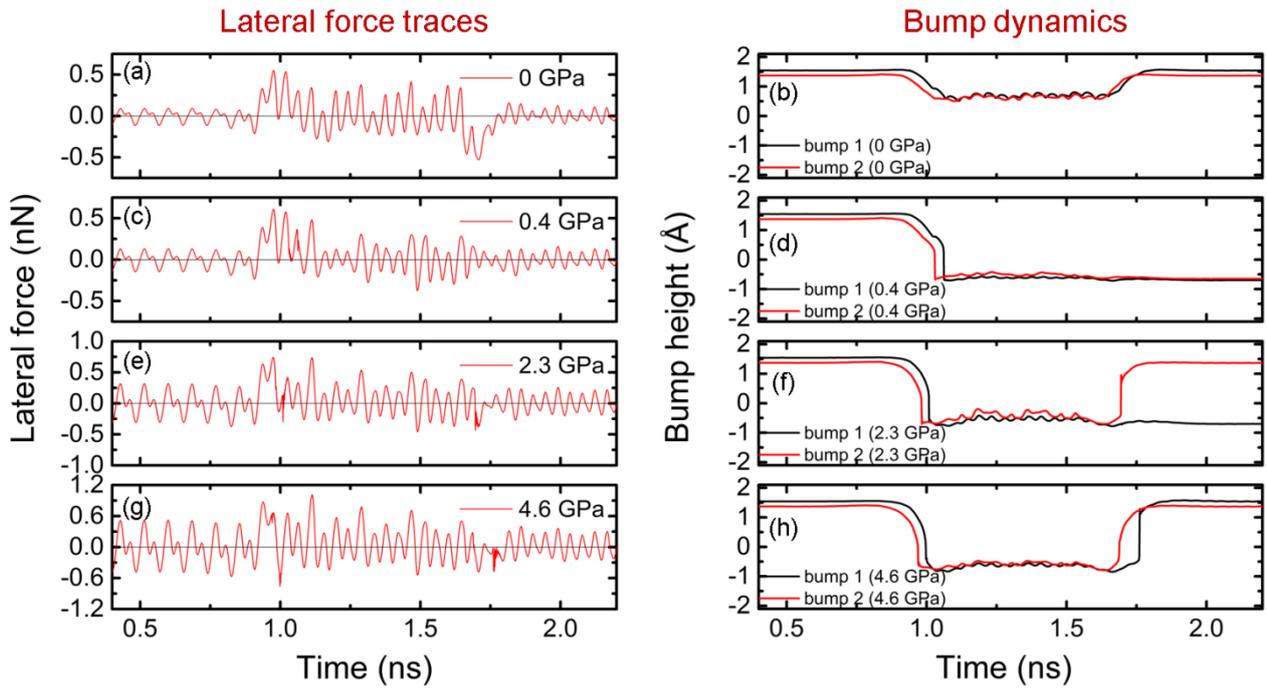

**Figure S16.** Lateral force traces and bump dynamics in the forward sliding direction for the $\theta = 13.9°$ GB. Panels (a), (c), (e), and (g) present the lateral force traces as a function of time for normal loads of 0, 0.4, 2.3, and 4.6 GPa, respectively. Panels (b), (d), (f), and (h) show the corresponding bump heights and bump vertical velocities variation profiles. Positive values indicate a resistive force.

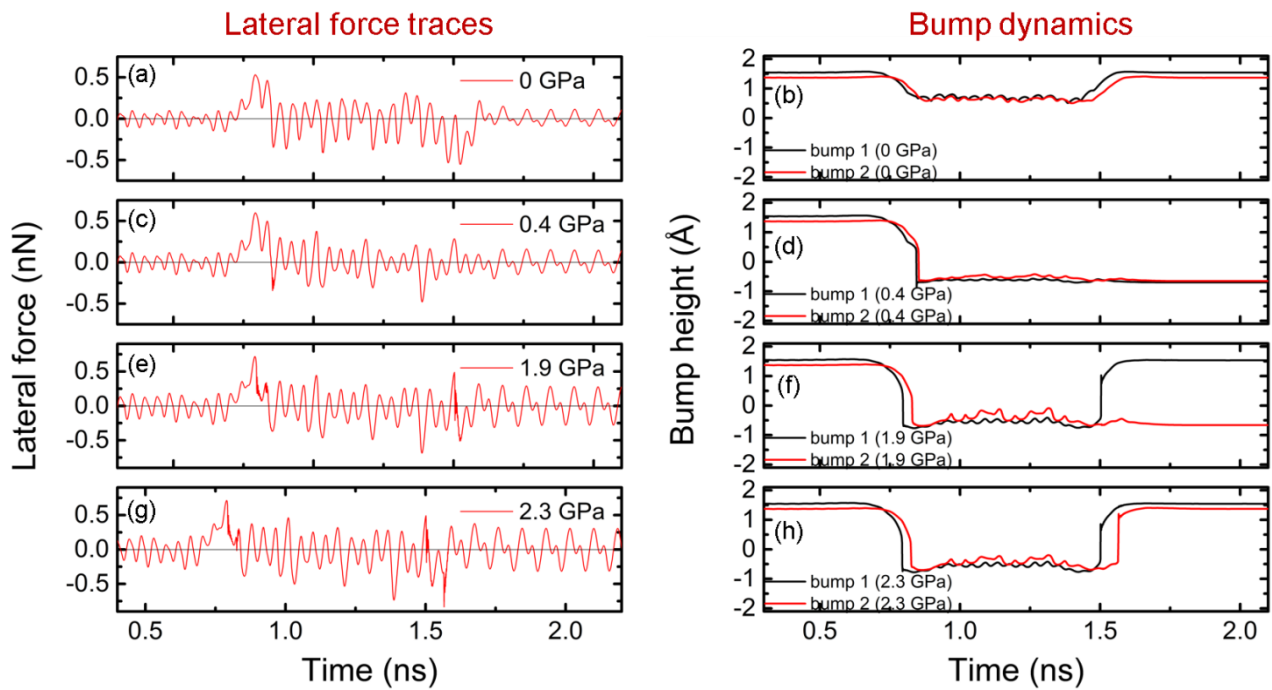

**Figure S17.** Lateral force traces and bump dynamics in the backward sliding direction for the $\theta = 13.9°$ GB. Panels (a), (c), (e), and (g) present the lateral force traces as a function of time for normal loads of 0, 0.4, 1.9 and 2.3 GPa, respectively. Panels (b), (d), (f), and (h) show the corresponding bump heights and bump vertical velocities variation profiles. Positive values indicate a resistive force.

## 12.3. Forward and Backward Sliding for the flat $\theta = 27.8°$ GB

The sliding simulation setup for the $\theta = 27.8°$ flat GB is shown in Figure S18. The range of the color scale is set to clearly demonstrate the small atomic height difference (~0.07 Å) between the two grains. The simulation protocol is same as that for the $\theta = 4.7°$ and 13.9° GBs. The lateral force traces at different normal loads in the forward (left column) and backward (right column) sliding directions are shown in Figure S19. In contrast with the corrugated GBs discussed above, the lateral force traces do not exhibit the typical signatures of bump buckling and unbuckling regardless of the applied normal load (see SI Movie 10). The various force trace patterns observed along the sliding path result mainly from the different registry matching between the TLG flake bottom layer and the various polycrystalline regions (Grain 1, Grain 2, and the GB). Notably, the forward and backward force traces are nearly inverted images of each other in this case.

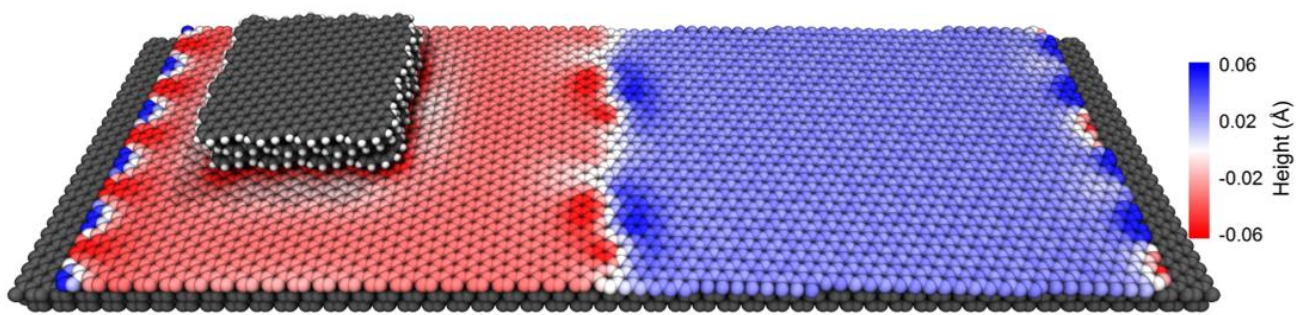

**Figure S18.** Sliding simulation setup for the $\theta = 27.8°$ GB. The grey and white spheres represent carbon and hydrogen atoms, respectively. Coloring of the polycrystalline atoms represents the atomic height with respect to the average height of the two grains (see color scale).

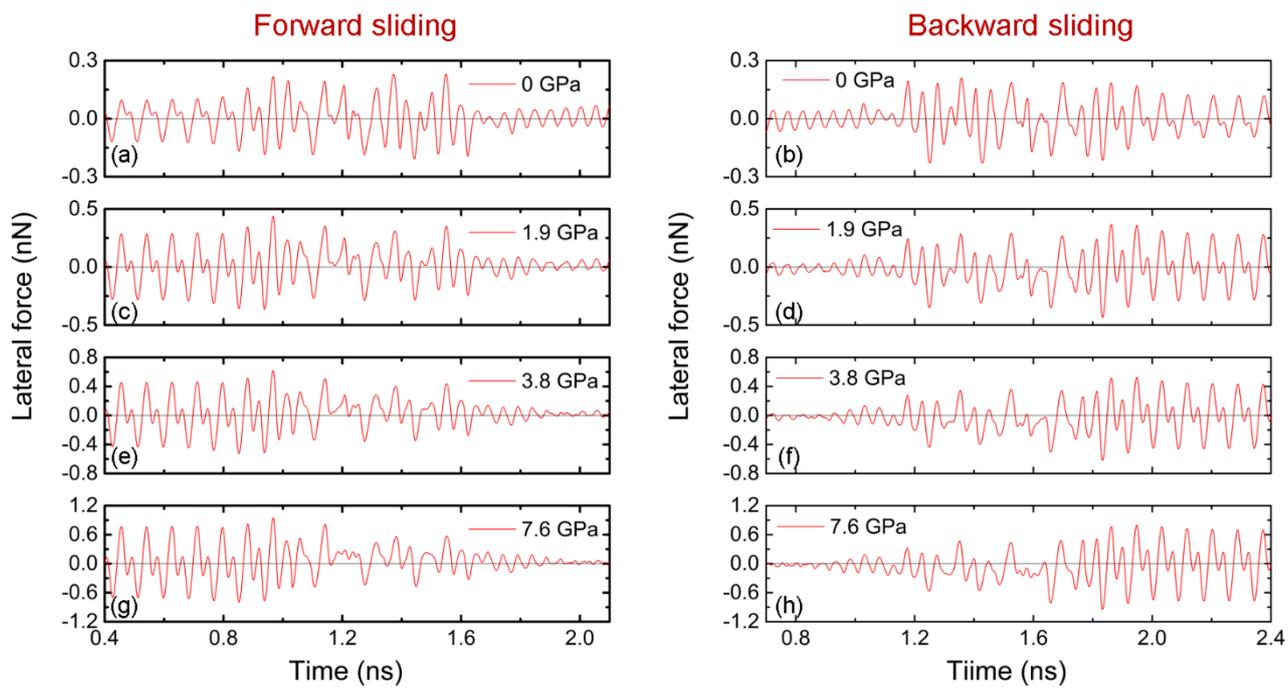

**Figure S19**. Lateral force traces along the forward (left panels) and backward (right panels) sliding direction for the $\theta =$ 27.8° GB at normal loads of (a), (b) 0; (c), (d) 1.9; (e), (f) 3.8; and (g), (h) 7.6 GPa.

## 12.4. Sliding Simulation for a Pristine Graphene Substrate

The sliding simulation setup for a pristine graphene substrate, shown in Figure S20a, is built by replacing the polycrystalline graphene layer with a periodic pristine graphene. The same simulation protocol for sliding on the polycrystalline graphene substrates was used. The sliding simulation for each normal load was conducted for 1 ns. As an example, the lateral force trace for a normal load of 0.8 GPa is shown in Figure S20b, showing well defined periodic behavior with the periodicity of the graphene lattice in armchair direction (4.26 Å). The friction force is calculated by averaging the lateral forces at steady-state over six force oscillation periods (e.g. the interval between the dash lines in Figure S20b), neglecting the initial transient dynamics.

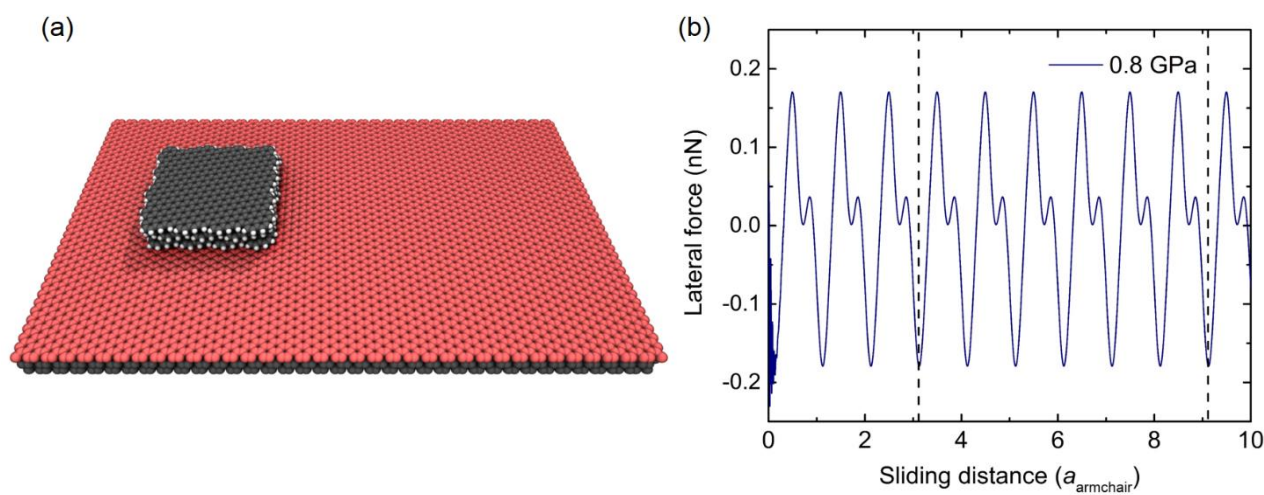

**Figure S20.** Simulations of friction over a pristine graphene substrate. (a) Simulation setup. The grey and white spheres represent carbon and hydrogen atoms, respectively. The top substrate layer is colored in red for clarity of the presentation. (b) Lateral force trace at a normal load of 0.8 GPa. The sliding distance is normalized by the periodicity of the graphene lattice in the armchair direction, $a_{armchair}$ = 4.26 Å. The dash lines mark the region for lateral force averaging.

## 13. Potential Energy Analysis

### 13.1. Comparison between Dynamic Simulations and Static Potential Energy Mapping

The static potential energy mapping of the flat GBs was performed by placing TLG flake at different positions along the sliding path and relaxing the system under an external normal load while freezing the lateral motion of the top TLG flake layer. These calculations provide a good representation of the potential energy profile along the sliding path. However, to obtain the fine details of the full energy profile, many calculations are required (see Figure S21). A more efficient way (in lack of massive parallelization capabilities) to obtain the energy profile is to follow the potential energy component during the dynamical sliding simulation. To verify that both methods indeed provide a similar potential energy profiles and that dynamical effects are negligible in this respect, we compared the energy profiles obtained via static and dynamic calculations at several representative normal loads. An example of one such comparison for a normal load of 3.8 GPa is given in Figure S21, showing excellent agreement between the static (full black squares) and dynamic (full blue triangles) results. Finally, we present also representative results of quasi-static sliding simulations over Grain 1 and the GB regions (full red circles), where the top layer of the TLG flake is repeatedly shifted followed by structural relaxation of the whole system keeping the lateral position of the top layer fixed. Here, as well, excellent agreement between the quasi-static results and the results of the dynamic simulations is found. Therefore, we conclude that it is safe to employ dynamic simulation results for the potential energy analysis.

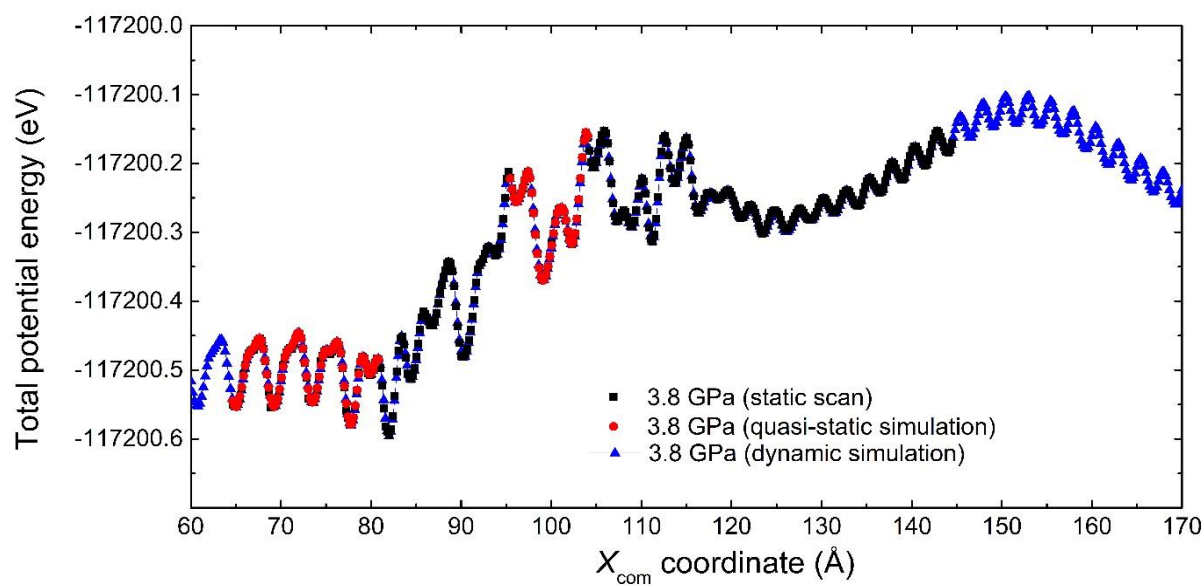

**Figure S21.** Comparisons of potential energy profiles obtained by the static scan (full black square), quasi-static calculations (full red circles), and dynamic simulations (full blue triangle), as a function of the center of mass position of the top layer of the flake for the flat $\theta = 27.8°$ GB at a normal load of 3.8 GPa.

## 13.2. Potential Energy Components for the Flake and the Interface of the $\theta = 27.8°$ GB

The potential energy profiles of the flake and the interaction energy between the flake and the substrate, calculated from dynamic simulations when crossing the GB, are shown in Figures S22-23, respectively. The potential profiles of the flake show minor differences between the two grains and are insensitive to the normal load. The interaction energy shows an energy difference of ~0.1 eV between the two grains, which is insensitive to the normal load, as well. Since both contributions do not vary with normal load, our discussion regarding the load dependence friction mechanism in the main text excluded these two terms and focused on the substrate elastic energy contribution.

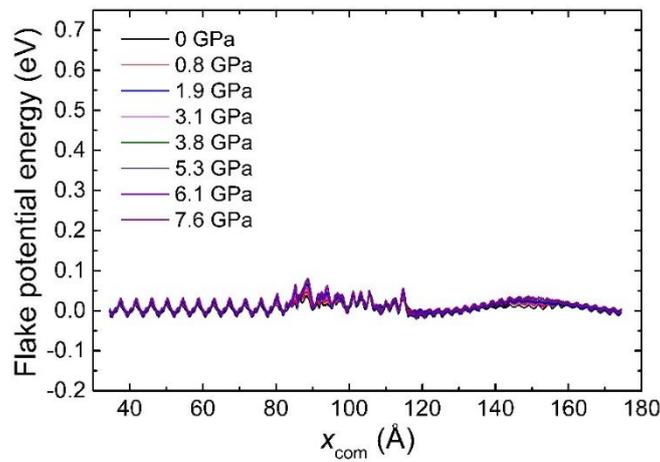

**Figure S22.** The potential energy profiles of the flake crossing the $\theta = 27.8°$ GB as a function of the center of mass position of the top layer of the sliding flake at different normal loads. For comparison purposes, the substrate potential energy under each normal load when the flake is positioned deep inside Grain 1 is set to zero.

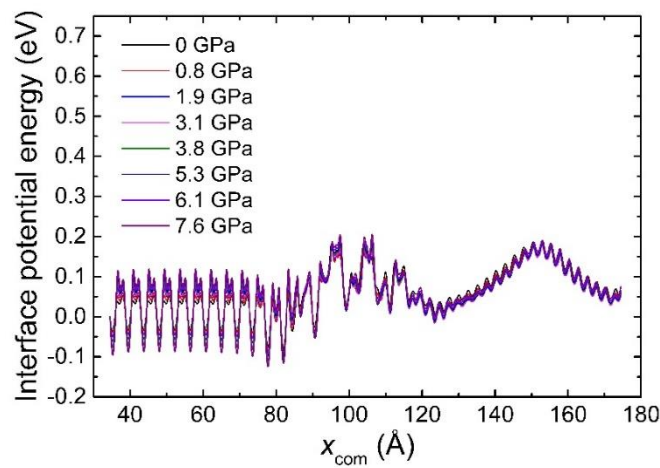

**Figure S23.** The interaction energy profiles of the flake-substrate interface for when crossing the $\theta = 27.8°$ GB as a function of the center of mass position of the top layer of the sliding flake at different normal loads. For comparison purposes, the substrate potential energy under each normal load when the flake is positioned deep inside Grain 1 is set to zero.

## 14. Energy Dissipation Analysis

The total energy dissipation contains two contributions: the dissipated kinetic energy pulses (as shown in Figure S12a and the stored potential (elastic) energy due to depressed or buckled bumps. The latter can be calculated from the potential energy profiles obtained by the NEB calculations illustrated in section 9. For the $\theta = 4.7°$ and $13.9°$ GBs, the total energy dissipation due to buckling, for forward sliding, backward sliding, and their average is shown in Figure S24. A steep increase of the dissipated energy is found when buckling occurs (at normal loads $\lesssim 0.4$ GPa for $\theta = 4.7°$ and 0 GPa for $\theta = 13.9°$) where kinetic energy is dissipated and elastic energy is stored in the buckled bump. As the normal load increases, the kinetic energy dissipated during a buckling event is gradually reduced and the stored potential (elastic) energy is released due to unbuckling processes, until finally the energy loss due to bump buckling becomes negligible at high normal loads. This nonlinear buckling energy loss behavior leads to the non-Amonton variation of the friction force shown in Figure 3c-e of the main text. Specifically, the buckling energy losses presented in Figure S24a,b account for the nonlinear parts of the friction stress for forward sliding and backward sliding (Figure 3c,d in the main text), respectively. The average buckling energy loss (Figure S24c) exhibits a similar behavior as that of the average friction stress (Figure 3e in the main text).

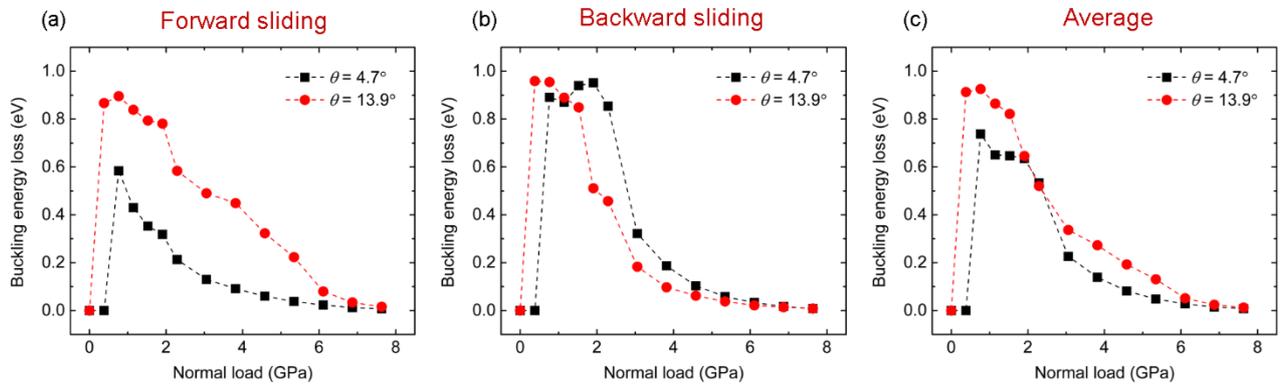

**Figure S24.** Buckling energy dissipation into kinetic and elastic components for (a) forward sliding, (b) backward sliding, and (c) their average over the $\theta = 4.7°$ (black squares) and $\theta = 13.9°$ (red circles) GBs as a function of normal load. For normal loads below the buckling load, the energy loss associated with bump deformation is negligible and hence set to zero.

## 15. Sensitivity Analysis of the Damping Rate Applied to the Top Layer of the Flake

In the dynamic simulations presented through this study, we applied damping to the vertical motion of the top layer of flake to suppress exaggerated vertical oscillations of the slider while crossing the GB. In experimental scenarios such oscillations are naturally suppressed by energy dissipation to the various components of the measurement setup. To this end, we adopt a damping coefficient of 1.0 ps$^{-1}$, same as that applied to all degrees of freedom of the middle layers of the substrate and the slider. To verify that this choice of damping coefficient has minor effect on the obtained results we repeated our simulations for the slider crossing the $\theta = 4.7°$ GB under a normal load of 0.8 GPa using a damping rate of 0.1 ps$^{-1}$. Figure S25 presents the vertical force acting on the top TLG flake layer clearly showing that the force fluctuations during the bump buckling and unbuckling events indeed decay slower with decreasing damping rate, as expected. Notably, as can be seen in Figure S26 the lateral force measured during this process is practically insensitive to the choice of vertical motion damping rate within this parameter range, typically used in friction simulations.

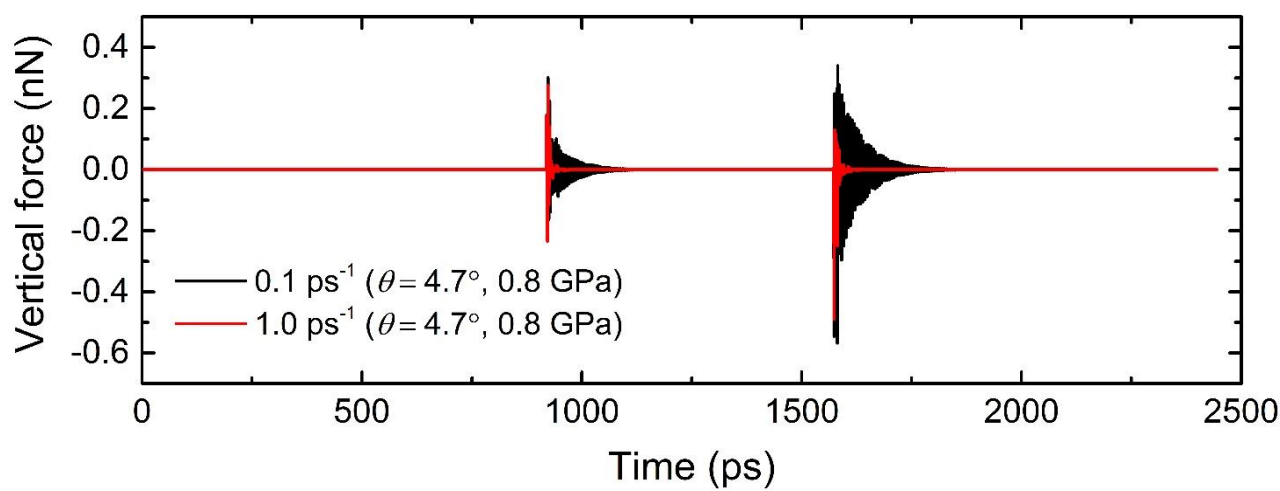

**Figure S25.** Vertical force traces acting on the top layer of the TLG flake sliding over the $\theta = 4.7°$ GB under a normal load of 0.8 GPa using two damping coefficients of 0.1 ps$^{-1}$ (black line) and 1.0 ps$^{-1}$ (red line).

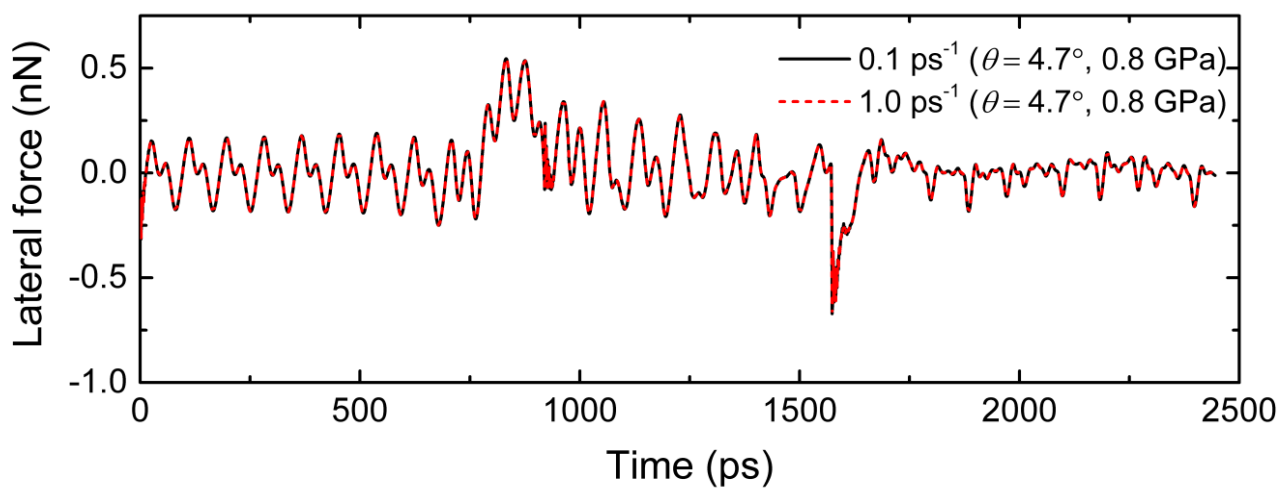

**Figure S26.** Lateral force traces acting on a TLG flake crossing the $\theta = 4.7°$ GB system under a normal load of 0.8 GPa using two damping coefficients of 0.1 ps$^{-1}$ (solid black line) and 1.0 ps$^{-1}$ (dashed red line).